\documentclass[twocolumn,english,prl,floatfix,twocolumn,showpacs]{revtex4-1}
\usepackage[T1]{fontenc}
\usepackage{ulem,xcolor}
\usepackage[latin9]{inputenc}
\usepackage{amsbsy}
\usepackage{amstext}
\usepackage{amsmath}
\usepackage{graphicx}
\usepackage{color}
\makeatletter

\@ifundefined{textcolor}{}
{%
 \definecolor{BLACK}{gray}{0}
 \definecolor{WHITE}{gray}{1}
 \definecolor{RED}{rgb}{1,0,0}
 \definecolor{GREEN}{rgb}{0,1,0}
 \definecolor{BLUE}{rgb}{0,0,1}
 \definecolor{CYAN}{cmyk}{1,0,0,0}
 \definecolor{MAGENTA}{cmyk}{0,1,0,0}
 \definecolor{YELLOW}{cmyk}{0,0,1,0}
}

\global\long\def\ket#1{\left| #1\right\rangle }
\global\long\def\bra#1{\left\langle #1 \right|}
\global\long\def\kket#1{\left\Vert #1\right\rangle }
\global\long\def\bbra#1{\left\langle #1\right\Vert }
\global\long\def\braket#1#2{\left\langle #1\right. \left| #2 \right\rangle }
\global\long\def\bbrakket#1#2{\left\langle #1\right. \left\Vert #2\right\rangle }
\global\long\def\av#1{\left\langle #1 \right\rangle }
\global\long\def\tr{\text{Tr}}
\global\long\def\im{\text{Im}}
\global\long\def\re{\text{Re}}

\newcommand{\teff}{T$_{\mbox{\tiny eff~}}$}

\newcommand{\be}{\begin{eqnarray}}
\newcommand{\ee}{\end{eqnarray}}\def\beq{\begin{equation}}\def\eeq{\end{equation}}

\usepackage{babel}
\makeatother

\begin{document}

\title{Steady-state dynamics and effective temperatures of quantum criticality in an open system}

\author{P. Ribeiro}
\affiliation{Russian Quantum Center, Novaya street 100 A, Skolkovo, Moscow area, 143025 Russia}
\affiliation{Centro de F\'\i sica das Interac\c c\~oes Fundamentais,
Instituto Superior T\'ecnico, Universidade de Lisboa,
Av. Rovisco Pais, 1049-001 Lisboa, Portugal }
\author{F. Zamani}
\affiliation{Max Planck Institute for Physics of Complex Systems, 01187 Dresden,
Germany}
\affiliation{Max Planck Institute for Chemical Physics of Solids, 01187 Dresden,
Germany}
\author{S. Kirchner}\email{stefan.kirchner@correlated-matter.com}
\affiliation{Center for Correlated Matter, Zhejiang University, Hangzhou,  Zhejiang 310058, China}

\begin{abstract}
We study the thermal and non-thermal steady state scaling functions and the steady-state dynamics of a  model of local quantum criticality. The model we consider, {\it i.e.} the pseudogap Kondo model, allows us to study the concept of effective temperatures near fully interacting as well as weak-coupling  
fixed points. In the vicinity of each fixed point we establish the existence of an effective temperature --different at each fixed point-- such that the equilibrium fluctuation-dissipation theorem is recovered. Most notably, steady-state scaling functions in terms of the effective temperatures coincide with the equilibrium scaling functions. This result extends to higher correlation functions as is explicitly demonstrated for the Kondo singlet strength. The non-linear charge transport is also studied and analyzed in terms of the effective temperature. 
\end{abstract}

\maketitle

\global\long\def\ket#1{\left| #1\right\rangle }
\global\long\def\bra#1{\left\langle #1 \right|}
\global\long\def\kket#1{\left\Vert #1\right\rangle }
\global\long\def\bbra#1{\left\langle #1\right\Vert }
\global\long\def\braket#1#2{\left\langle #1\right. \left| #2 \right\rangle }
\global\long\def\bbrakket#1#2{\left\langle #1\right. \left\Vert #2\right\rangle }
\global\long\def\av#1{\left\langle #1 \right\rangle }
\global\long\def\tr{\text{tr}}
\global\long\def\Tr{\text{Tr}}
\global\long\def\pd{\partial}
\global\long\def\im{\text{Im}}
\global\long\def\re{\text{Re}}
\global\long\def\sgn{\text{sgn}}
\global\long\def\Det{\text{Det}}
\global\long\def\abs#1{\left|#1\right|}
\global\long\def\up{\uparrow} 
\global\long\def\down{\downarrow}
\global\long\def\k{\mathbf{k}}
\global\long\def\wks{\mathbf{\omega k}\sigma}
\global\long\def\vc#1{\mathbf{#1}}
\global\long\def\bs#1{\boldsymbol{#1}}
%
The interest in understanding the dynamics of strongly correlated systems beyond the linear response regime has in recent years grown tremendously.\\
The quantum dynamics  in adiabatically  isolated optical traps has been successfully modeled using powerful numerical schemes~\cite{Eckstein2010,  Arrigoni2013}. In open systems mainly diagrammatic techniques on the Schwinger-Keldysh contour have been employed. For nanostructured systems  several techniques exist to describe the ensuing out-of-equilibrium properties. These approaches, however,  are either perturbative in nature~\cite{Munoz.13}, centered around high temperatures and short times~\cite{Werner2009,   Gull2011,  Cohen2013,  Aoki2014,  Schiro2010}, or approximate the continuous baths by discrete Wilson chains~\cite{Rosch.12,Nghiem.14,Nghiem.a.14}.
The situation might be simpler for non-linear  dynamics that arises in the vicinity of a quantum critical point (QCP),  where a vanishing energy scale leads to scaling and universality~\cite{  Mitra.06, Diehl2008,  Hogan.08, Chung2009,  Kirchner.09,  Takei2010, Ribeiro2013b, Sieberer2013}.\\  
For the dynamics near classical continuous phase transitions a well-established theoretical framework exists, tying the dynamics  to the  statics and the conserved quantities~\cite{Hohenberg.77}.  
In addition, the concept of effective temperature (\teff) was established as an
useful notion for the  relaxational dynamics of classical critical systems~\cite{Hohenberg1989, Cugliandolo.97,Calabrese.04, Bonart.12}, although it appears somewhat less useful for fully interacting critical points~\cite{Calabrese.04}.
\teff is commonly defined by extending the equilibrium fluctuation-dissipation theorem to the non-linear regime.
The existence of effective temperatures in quantum  systems was recently investigated~\cite{Cugliandolo.11,Mitra2005,Kirchner2010,Caso2011, Ribeiro2013b}. For a recent review see~\cite{Cugliandolo.11}.
In comparison to classical criticality, at a QCP, dynamics already enters at the equilibrium level. For a QCP that can be described by a Ginzburg-Landau-Wilson  functional in elevated dimensions, it was found that the voltage-driven transition is in the universality class of the associated thermal classical model with voltage acting as \teff~\cite{Mitra.06}.  Unconventional QCPs in contrast are not described solely in terms of an order parameter functional~\cite{Gegenwart.08,Zhu.06}.

\begin{figure}
\centering{}
\includegraphics[width=0.8\linewidth]{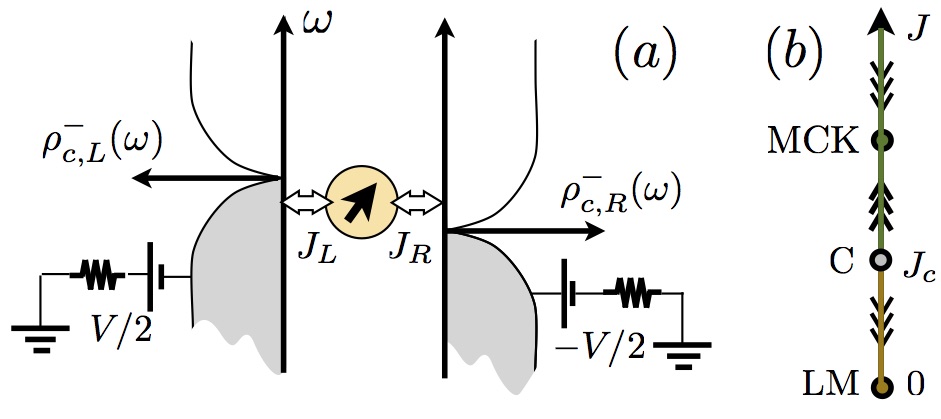}
\protect\caption{\label{figure1}(a) Sketch of the model: a spin interacts with two fermionic leads which are characterized by their
respective density of states $\rho_{c,L/R}^{-}\left(\omega\right)$ and chemical potential $\mu_{L/R}$. 
(b) Phase diagram  of the 
multichannel pKM with gap exponent $r<r_{\mbox{\tiny max}}$: A QCP (C) separates the multichannel Kondo fixed point (MCK) from the (weak-coupling) local moment fixed point (LM).} 
\end{figure}

In this letter we address the following general questions within a model system of unconventional quantum criticality:
Is the existence of \teff tied to dynamical (or $\omega/T$-)scaling? Does \teff have meaning for higher correlation functions? How unique is \teff at a given fixed point once boundary conditions have been specified?
Can critical scaling functions be expressed through \teff and if so, how do these scaling functions 
relate to the equilibrium scaling functions?
The model system is the pseudogap Kondo (pKM) model that describes a quantum spin anti-ferromagnetically coupled to a conduction electron bath possessing a pseudogap near its Fermi energy, characterized by a powerlaw exponent. Depending on the coupling strength, the quantum spin is either screened or remains free in the zero temperature ($T$) limit. The two phases are separated by a critical point dispaying critical Kondo destruction, see Fig.\ref{figure1}.
The pKM  has been invoked to describe non-magnetic impurities in the cuprate superconductors~\cite{Vojta.02} and point-defects in graphene~\cite{Chen2011}. It underlies the pseudogap free moment phases occurring in certain disordered metals~\cite{Zhuravlev.07} and can also be realized in
double quantum-dot systems~\cite{Diasdasilva.06}. The quantum critical properties of the pKM in equilibrium have been addressed in~\cite{Withoff1990,Bulla.97,Buxton.98,Logan.00,Vojta2001,Ingersent.02,Glossop.03,Glossop.05,Fritz2006,Glossop.11,Fritz.13}.
Our main findings are that the steady state dynamic spin susceptibility, the conductance, and the Kondo-singlet strength, a 4-point correlator,  reproduce their equilibrium behavior in the scaling regimes of the fixed points of the model when expressed in terms of a fixed-point specific \teff.\\
{\it The model.}
We consider a pKM  with a density of states that vanishes in a power-law fashion with exponent $0\leq r \leq 1$ at their respective Fermi level, $\rho_{c,l}^{-}(\omega)\sim |\omega|^r\Theta(D-|\omega|)$, with half-bandwidth $D$. Here, $l=L,R$ labels the two leads, see Fig.\ref{figure1}(a). 
In the multichannel version of the model the spin degree of freedom ($\bs S$) is generalized from $SU(2)$ to $SU(N)$ and the
fermionic excitations ($c$) of the leads transform under the fundamental
representation of $SU(N)\times SU(M)$ with $N$ spin and $M$ charge
channels.
At $T=0$ and $r<r_{\mbox{\tiny max}}<1$, a critical point  (C) separates a multichannel Kondo (MCK)-screened phase from a local moment (LM) phase at a critical value $J_c$ of the exchange coupling $J>0$, see Fig.\ref{figure1}(b).
The characterization of the phases and the leading power law exponents of observables of the pKM have been obtained by perturbative RG, large-$N$ methods, and NRG~\cite{Withoff1990,Buxton.98,Logan.00,Fritz2006, Bulla.97,Buxton.98,Ingersent.02}. 
Within the large-$N$
approach, at $T=0$, scaling arguments are able to predict the critical
exponents of dynamical observables~\cite{Vojta2001,Zamani}.
Non-equilibrium steady-states (NESS) are obtained by applying a time-independent bias voltage $V=(\mu_L-\mu_R)/\abs{e}$, where $\mu_l$  is the chemical potential of lead $l$, see Fig.~\ref{figure1}(a). As $T$  characterizes the fermionic reservoirs, it remains well-defined even for $V\neq0$.\\
A similar setup has been considered in a  perturbative RG-like study  adapted to the NESS condition~\cite{Chung2012}.
This model has also been invoked in a  variational study of the dynamics following a local
quench where it was found that quenches in the Kondo phase thermalize while this in not the case for quenches across the QCP into the LM regime~\cite{Schiro2012a}.
%
%
%
The system is described
by the Hamiltonian
\begin{eqnarray}
H & = & \sum_{p,\alpha\sigma l}\varepsilon_{pl}c_{p\alpha\sigma l}^{\dagger}c_{p\alpha\sigma l}+\frac{1}{N}\sum_{ll'}\sum_{\alpha}J_{ll'}\mathbf{S}.\mathbf{s}_{\alpha;ll'},
\label{eq:Hamiltonian_PGK}
\end{eqnarray}
where $\sigma=1,\ldots,N$ and $\alpha=1,\ldots,M$ are, respectively, the $SU(N)$-spin
and $SU(M)$-channel indices, $l$ labels the  leads
and $p$ is a momentum index. The co-tunneling term \cite{Kaminski2000}
in Eq.~(\ref{eq:Hamiltonian_PGK}) contains the local operators $\mathbf{s}_{\alpha;ll'}^{i}=\frac{1}{n_{c}}\sum_{pp'\sigma\sigma'}c_{p\alpha\sigma l}^{\dagger}t_{\sigma\sigma'}^{i}c_{p'\alpha\sigma'l'}^{\dagger}$
with $t$ the fundamental representation of $SU(N)$ and $n_{c}$
is the number of fermionic single-particle states. 
In a totally anti-symmetric representation, one can decompose the spin operator as
$S_{\sigma \sigma'}=f^\dagger_{\sigma} f^{}_{\sigma'}-q \delta_{\sigma \sigma'}$, where $q$ is subject to the constraint 
 $\hat{Q}=\sum_{\sigma}f_{\sigma}^{\dagger}f_{\sigma}=qN$ and
the $f_{\sigma}^{\dagger},f_{\sigma'}^{}$ obey fermionic commutation relations.

We employ a dynamical large-N limit~\cite{Parcollet1998,Vojta2001}, suitably generalized to the Keldysh contour~\cite{Kirchner.09,Ribeiro2013b} 
while keeping  $q=\frac{Q}{N}$ and $\kappa=M/N$ constant.
 This results in
\begin{eqnarray}
\Sigma_{B}^{>,<}\left(t\right) & = & iG_{f}^{>,<}\left(t\right)G_{c}^{<,>}\left(-t\right)\label{eq:Self_1}\\
\Sigma_{f}^{>,<}\left(t\right) & = & -i\kappa G_{B}^{>,<}\left(t\right)G_{c}^{>,<}\left(t\right)
 \label{eq:Self_2}\\
-iG_{f}^{<}\left(0\right) & = & q,
\end{eqnarray}
where $G_f$ is the pseudofermion propagator and $G_B$ is the propagator of a bosonic Hubbard-Stratonovich decoupling field.
$\Sigma_{f}$ ($\Sigma_{B}$) is the proper selfenergy of $G_f$ ($G_B$) and is related to it via the Dyson equation~\cite{suppl}.
We assume that the exchange interaction originates from an Anderson-type model via a Schrieffer-Wolff transformation, so that
a single
coupling constant $J=J_{L}+J_{R}$ emerges~\cite{suppl}.\\
For details on the numerics see ~\cite{suppl}.
In equilibrium, our approach yields dynamical scaling functions that coincide with those obtained from quantum Monte-Carlo~\cite{Glossop.11}.  
%
%
%
%
\begin{figure}
\centering{}
\includegraphics[width=1\linewidth]{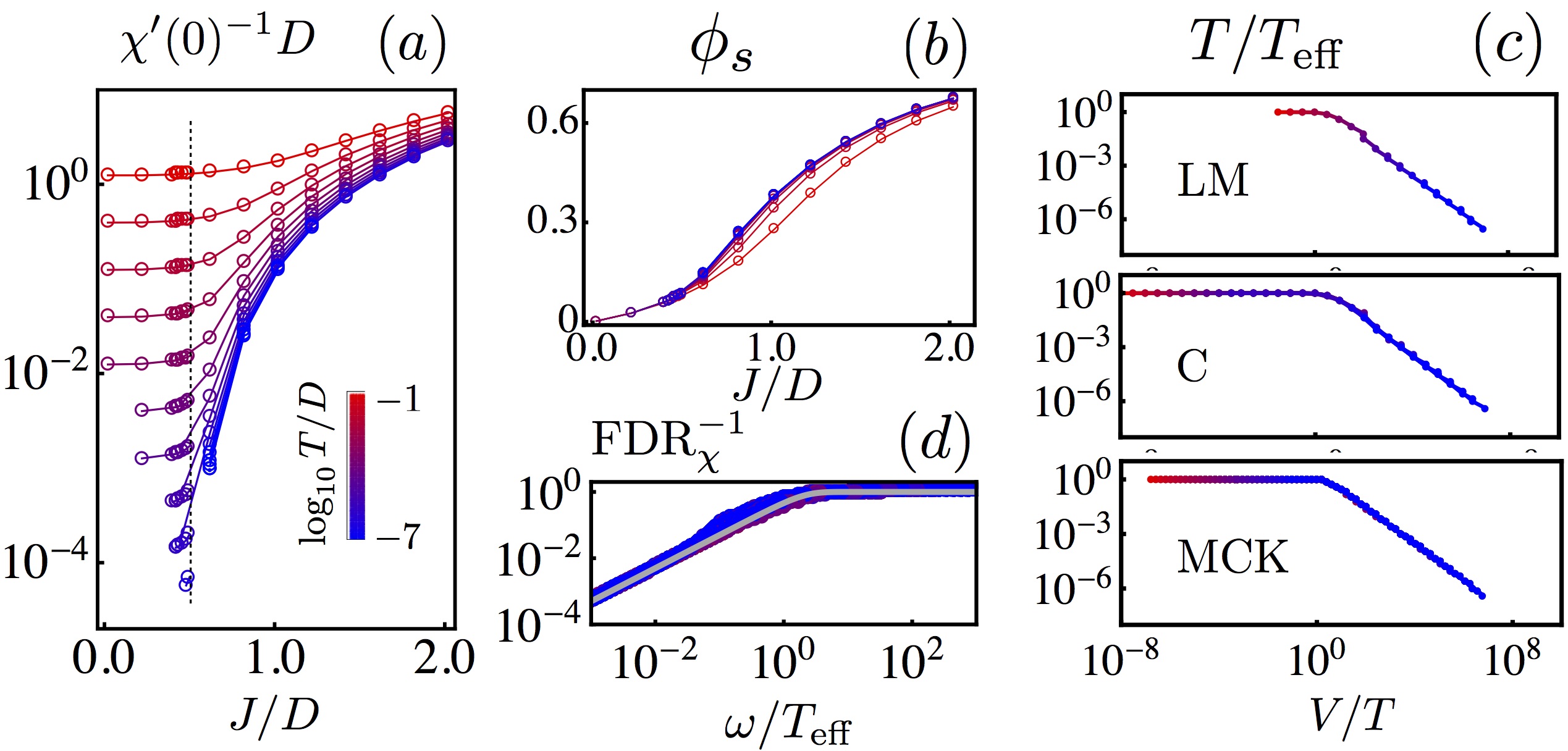}
\protect\caption{\label{fig:equilibrium}(a) $\chi'\left(0\right)^{-1}$vs
$J$ for different $T$. 
(b) $\phi_s$ vs $J$  for different $T$. The $T=0$ curve is approached from below in the MCK and from above
in the LM phases.
(c)  Scaling $T/T_{\mbox{\tiny eff}}$ vs $V/T$ at fixed points LM, C, and MCK: $T_{\mbox{\tiny eff}}\sim V$ for $V\gg T$. 
(d) $\text{FDR}^{-1}_{\chi}(\omega)$ vs $\omega/T_\text{eff}$ near fixed point C, shown for $V/D = 10^{-2}, 10^{-3}, 10^{-4}, 10^{-5}, 10^{-6}$.  The grey line is $\text{FDR}^{-1}(\omega)= \tanh(\beta \omega /2 )$.
} 
\end{figure}

{\it Observables.}
A possible order parameter for the transition from the overscreened Kondo to local-moment phase is given by 
$\lim_{T\rightarrow 0} T \chi(\omega=0,T)$, where $\chi(\omega,T)$ is the Fourier transform of the local (impurity)  spin-spin correlation function $\chi(t-t')$, see Fig.~\ref{fig:equilibrium}(a).
We work on the Keldysh contour where the lesser and greater components are defined in the usual way as 
 $\chi^{>}\left(t-t'\right)=-i\frac{1}{N}\sum_{a}\av{S^{a}\left(t\right)S^{a}\left(t'\right)}$ with $t\in\gamma_{\leftarrow}$ and $t'\in\gamma_{\rightarrow}$
and $\chi^{<}\left(t-t'\right)=-i\frac{1}{N}\sum_{a}\av{S^{a}\left(t'\right)S^{a}\left(t\right)}$, with $t\in\gamma_{\rightarrow}$ and $t'\in\gamma_{\leftarrow}$ so that
$\chi^{R}\left(t\right)=\Theta\left(t\right)\left[\chi^{>}\left(t\right)-\chi^{<}\left(t\right)\right]$ and $\chi^A=\chi^R+\chi^<-\chi^>$.
Here, $\gamma_{\rightarrow(\leftarrow)}$ is the forward (backward) branch of the Keldysh contour, respectively.

We also consider the  ``singlet-strength'' $\phi_{s}$, defined through the Kondo term contribution to the total energy of the system as
$\frac{1}{N}\sum_{ll'}\sum_{c}J_{ll'}\av{\mathbf{S}.\mathbf{s}_{c,ll'}}=-J\kappa\left(\frac{N^{2}-1}{N}\right)\phi_{s}$~\cite{Werner.12}.
$\phi_{s}$ is a dimensionless quantity, which possesses
a well-defined large-$N$ limit and  quantifies the degree of singlet formation. 
In terms of the fermionic fields, it can be written as the  local-in-time limit of a  4-point correlator~\cite{suppl}. 
Its equilibrium properties  will be discussed below.
The steady state charge current passing through each channel is
$\mathcal{J}_{\text{P}}=-\partial_{t}\av{\hat{N}_{L}\left(t\right)}/M$,
where $\hat{N}_{L}=\sum_{p\alpha\sigma}c_{p\alpha\sigma L}^{\dagger}c^{}_{p\alpha\sigma L}$
is the number of particles in the left lead. 
The out-of-equilibrium conditions
considered here respect particle-hole symmetry which implies a vanishing energy current.

Throughout the paper we set $\kappa=0.3$, $r=0.2$, and $q=1/2$. This results in $r_{\mbox{\tiny max}}=0.412(4)$.
Our choice of values for $\kappa$ and $r$  ensures a finite static spin susceptibility
$\chi'\left(\omega=0\right)$ within the MCK phase as $T\rightarrow 0$. We denote the real (imaginary) part of $\chi^R(\omega)$ by
$\chi'$ ($\chi''$).
\begin{figure}
\centering{}\includegraphics[width=1\linewidth]{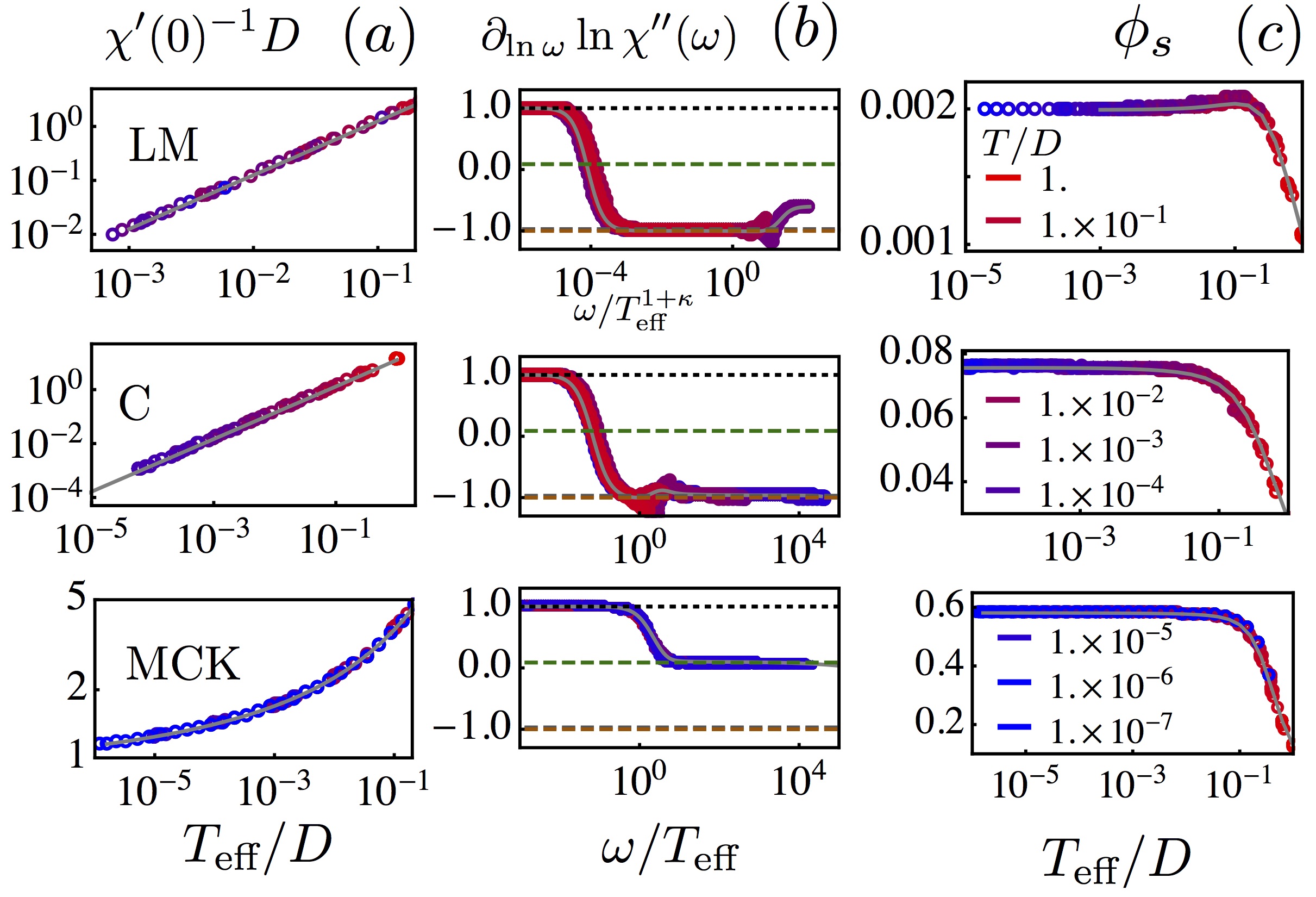}\protect\caption{\label{fig:effectiveT}
Scaling of observables with $T_{\text{eff}}$ at different fixed points for the values of $V$ as in Fig.~\ref{fig:equilibrium}(d):
(a) $\chi'\left(0\right)^{-1}$vs
$T_{\text{eff}}$; (b) $\protect\pd_{\ln\omega}\ln\chi''\left(\omega\right)$
vs $\omega/T_{\text{eff}}$; (c)  $\phi_{s}$ vs $T_{\text{eff}}$.
For each fixed point, the equilibrium scaling form (grey curves)
is compared with the same quantity under non-equilibrium conditions
and $T$ substituted by $T_{\text{eff}}$.} 
\end{figure}
{\it Thermal steady-state}.
The equilibrium ($V=0$) behavior of $\chi'\left(\omega=0,T\right)$ in the relaxational regime ($\omega\ll T$) near the MCK, C, and LM fixed points is shown in
Fig.\ref{fig:equilibrium}-(a).
For $J< J_{c}\simeq0.44 D $, {\it i.e.} in the LM phase, one observes
Curie-like behavior at lowest temperatures $\chi'\left(\omega=0,T\right)\propto T^{-1}$.
In the MCK phase ($J> J_{c}$ and with our choices of $\kappa$ and $r$), the $T=0$ susceptibility remains finite. 
The grey lines in Figs.\ref{fig:effectiveT}-(b) 
show the scaling plots of the logarithmic derivative of  $\chi'' \left(\omega \right)$ for different values of the temperature, {\it i.e.} $\pd_{\ln\omega}\ln\chi''\left(\omega\right)$ for
the different fixed points. Note that $\pd_{\ln\omega}\ln\chi''\left(\omega\right)\simeq\alpha_{\chi}$
within the scaling region where $\chi''\left(\omega\right)\propto\abs{\omega}^{\alpha_{\chi}}$.
The values of $\alpha_{\chi}$ in the quantum coherent regime ($\omega/T\gg1$)
agree with those obtained analytically
from a $T=0$ scaling ansatz \cite{Zamani} for
the MCK ($\alpha_{\chi}\simeq0.087$) and C ($\alpha_{\chi}=-0.97$)
fixed points. 
These results are compatible with a dynamical scaling form $\chi''\left(\omega,T\right)=T^{\alpha_{\chi}}\Phi\left(\omega/T\right)$,
in terms of an universal scaling function $\Phi\left(x\right)$ possessing
asymptotic values $\Phi\left(x\right)\propto x$ for $x\ll1$ and
$\Phi\left(x\right)\propto x^{\alpha_{\chi}}$ for $x\gg1$. Thus,
the scaling properties
are in line with dynamical $\omega/T$-scaling for the C and MCK fixed
points. For the LM fixed point we find $\alpha_{\chi}=-1$ and a scaling
form $\chi''\left(\omega\right)=T^{\alpha_{\chi}}\widetilde{\Phi}\left(\omega/T^{1+\kappa}\right)$, indicative of a weak-coupling 
fixed point and absence of hyperscaling.
These results will be further addressed elsewhere~\cite{Zamani}.\\
The singlet-strength $\phi_{s}$ vs.~$J$ at different $T$ and at $V=0$ is shown in
Fig.\ref{fig:equilibrium}-(b). 
The numerical data at $T\neq0$ suggest that $\phi_{s}\left(J,T=0\right)$
is a continuous function of $J$.  At the C fixed point we find that  $\phi_{s}\left(J,T\right)$ as a function of $J$
crosses for different values of $T$ (for sufficiently
low $T$).\\
{\it Non-thermal steady-states}.
We consider a non-equilibrium setup where the two leads, initially
decoupled from the impurity (for $t<t_{0}$), are held at chemical
potentials $\mu_{L}=-\mu_{R}=\abs eV/2$ ($\abs e=1$ in the following).
 At $t=t_{0}$ the coupling between the leads and
the impurity is turned on. A steady-state is reached by  
sending $t_{0}\to-\infty$ so that any transient behavior will already have faded away at (finite) time $t$. 
The NESS fluctuation-dissipation ratio (FDR)  for a dynamical
observable $A(t,t')=A(t-t')$ is defined through $\text{FDR}{}_{A}(\omega)=[A^{>}(\omega)+A^{<}(\omega)]/[A^{>}(\omega)-A^{<}(\omega)]$,
 where $A^{>/<}$ are the Fourier transforms of the greater/lesser
components of $A$.
At equilibrium, the fluctuation-dissipation
theorem implies $\text{FDR}_{A}(\omega)=\tanh\left(\beta\omega/2\right)^{\zeta}$
uniquely (with $\zeta=\pm1$ for fermionic (+) and bosonic (-) operators).
For a generic out-of-equilibrium system, the functional form of the
FDR differs from the equilibrium one. A frequency-dependent
\textquotedblleft effective temperature\textquotedblright , $1/\beta_{\text{eff}}^{A}\left(\omega\right)$,
for the observable $A$ can be defined by requiring that 
$\tanh\left[\beta_{\text{eff}}^{A}\left(\omega\right)\omega/2\right]^{\zeta}=\text{FDR}_{A}(\omega)$
\cite{Foini2011,Kirchner2010}.
Following Refs.~\cite{Hohenberg1989,Mitra2005,Ribeiro2013b}
we define $T_{\mbox{\tiny eff}}$ via  $\text{FDR}_{\chi}$
through
its asymptotic low-frequency behavior $T_{\text{eff}}^{-1}=\lim_{\omega\to0}\beta_{\text{eff}}^{\chi}\left(\omega\right)$. 
In equilibrium  $T_{\mbox{\tiny eff}}=T$. 
On the other hand, a linear-in-$V$ decoherence rate in the non-equilibrium relaxational regime near an interacting QCP is signaled by $\omega/V$-scaling~\cite{Kirchner.09}.
In this case and at $T=0$  one expects $T_{\mbox{\tiny eff}}=c V$, where $c$ characterizes the underlying fixed point.
We thus analyze  $T/T_{\mbox{\tiny eff}}$ vs $V/ T$.
Fig. \ref{fig:equilibrium}-(c)  shows the resulting $T/T_{\text{eff}}$ as a function
of $V/T$ for the different fixed points computed for
different values of $V$ and $T$. In the non-linear regime, the scaling collapse for $T/T_{\mbox{\tiny eff}}$ implies $T_{\mbox{\tiny eff}}=c V$, where $1/c$ is the amplitude of the scaling curve in the non-linear regime. A comparison of $\text{FDR}_{\chi}^{-1}$ with the equilibrium result for fixed point (C) is shown in Fig.\ref{fig:equilibrium}-(d). Even for the LM fixed point, where hyperscaling is violated, $T_{\mbox{\tiny eff}} \sim V$ holds for $V\gg T$, see Fig. \ref{fig:equilibrium}-(c), top panel. It is however important to realize that the properties we see in terms of \teff are  a property of the flow towards the LM fixed point. 
\begin{figure}[t!]
\centering{}\includegraphics[width=1\linewidth]{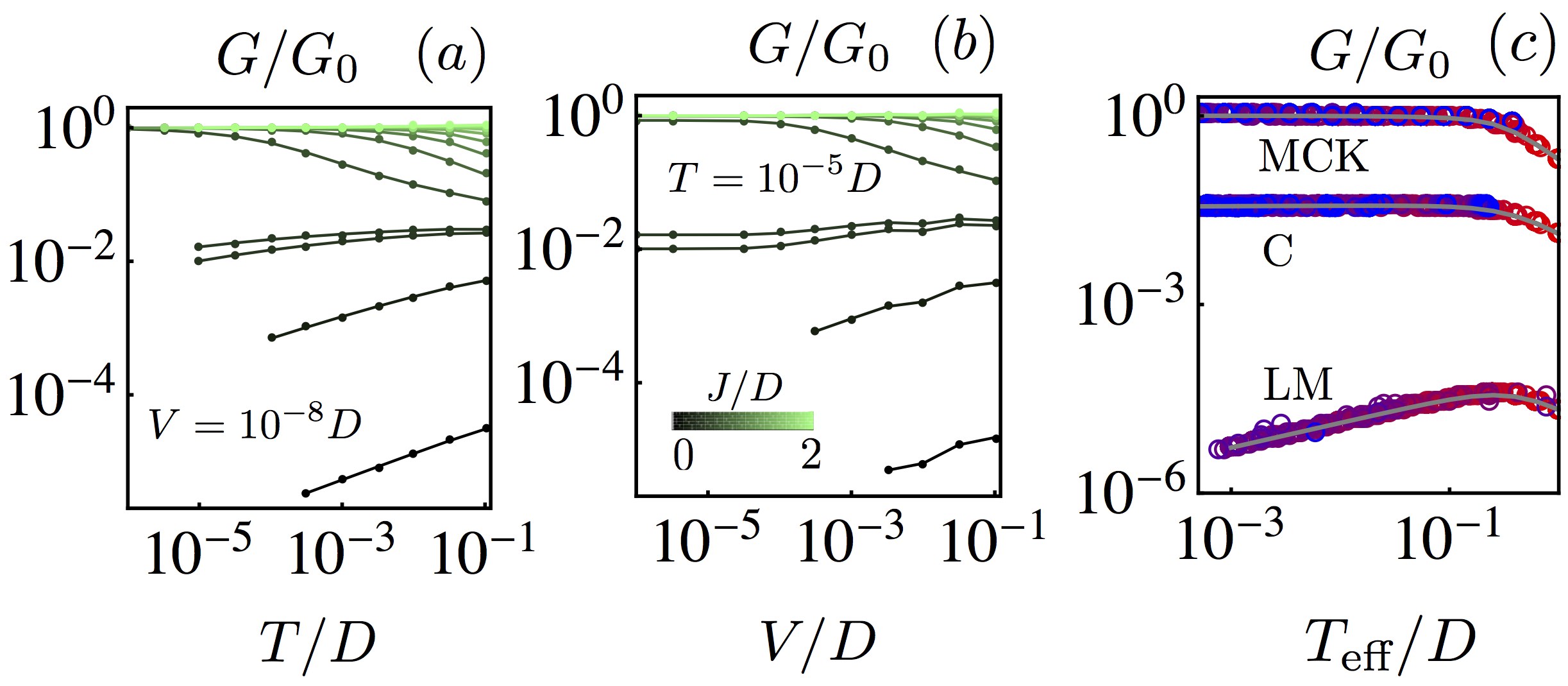}\protect\caption{\label{fig:transport}
Conductance $G$ normalised to the $\text{MCK}$ fixed point conductance $2\pi G_0 = 0.415$. 
(a) $G(T)$ vs T computed for the lowest non-zero value of $V$ at
different values of $J$ (see color coding). (b) $G$ vs $V$ for fixed $T$.
(c)  $G=\mathcal{J}_{\text{P}}/V$ vs $T_{\text{eff}}$ at different fixed points. The equilibrium form is given by the grey curves.}
\end{figure}
Far from equilibrium and outside any scaling regime, $\chi$ is a function of $\omega$, $T$, and $V$ but  near a fixed point $\chi(\omega,T,V)$ develops a scaling form in terms of a combination of $\omega$, $T$, and $V$. 
This then raises the question how $T_{\text{eff}}$ enters the scaling function and leads us to a remarkable result, see Fig. \ref{fig:effectiveT}-(b)-(c): The non-thermal steady-state scaling function of $\chi=\chi(\omega,T,V)$
when scaled in terms of \teff recovers the equilibrium scaling function of that particular fixed point with \teff replacing $T$. This not only turns out to be true for $\chi$ at each of the fixed points of the model but also holds for $\phi_s$, a higher-order correlation function. We first consider the static susceptibility.
Fig.~\ref{fig:effectiveT}-(a)
shows the equilibrium scaling forms of $\chi'\left(0\right)^{-1}$
as a function of $T_{\text{eff}}$ for different values of $T$
and $V$ for the LM, C and MCK fixed points. The color
coding reflects the values of  $T$ of the system.
The equilibrium form (grey lines) is recovered even for $T_{\text{eff}}/T\gg1$. 

A similar result can be obtained at finite $\omega$:
Fig.~\ref{fig:effectiveT}-(b) shows the log-derivative $\pd_{\ln\omega}\ln\chi''\left(\omega\right)$
as a function of $\omega/T_{\text{eff}}$ for different values of $T$ and $V$ for the LM, C and MCK fixed points.
These should be compared with the equilibrium results, the underlying grey lines:
The equilibrium scaling form is recovered by replacing $T$ by $T_{\text{eff}}$, both for $\omega \ll T_{\text{eff}}$ and $\omega \gg T_{\text{eff}}$  \footnote{Similar results hold for the Keldysh component of $\chi$.}.
Note that $T_{\text{eff}}$ is defined from the FDR of $\chi$
in the limit $\omega/T\to0$. Therefore, the fact that the equilibrium
scaling forms of $\chi'\left(0\right)$ and $\chi''\left(\omega\right)$
are reproduced for $T_{\text{eff}}/T\gg1$ and $\omega/T_{\text{eff}}\gg1$,
respectively, is remarkable. 
Fig. \ref{fig:effectiveT}-(c) depicts $\phi_{s}$ as a function of
$T_{\text{eff}}$ for different values of $T$ and 
$V$. Again, the equilibrium scaling behavior (gray curves) is reproduced.

Unlike $\chi$ and $\phi_s$, the conductance $G$ depends on both pseudoparticle propagators $G_f$ and $G_B$.
One thus may wonder if \teff can have any meaning  for $G$.
In Figs. \ref{fig:transport}-(a,b) we show the conductance
per channel $G=\mathcal{J}_{\text{P}}/V$ vs $T$ and $V$ respectively. 
In the linear response regime $V,T\ll T_{K}\left(J\right)$ of the MCK
phase, the current is proportional to the applied voltage $\mathcal{J}_{\text{P}}=G_{0}V$.
Outside of the scaling regime, i.e. for $V,T\gg T_{K}\left(J\right)$,
$G$ drops rapidly as $V$ or $T$ increase. The linear and non-linear current-voltage characteristics display power-law behavior as $T,V\to0$~\cite{Ribeiro2013b,Kirchner.09}.
Near C, {\it i.e.} for $J=J_{c}$, the relation
between $\mathcal{J}_{\text{P}}$ and $V$ is still linear, ($\mathcal{J}_{\text{P}}=G_{c}V$),
however the critical conductance $G_{c}$ is
much smaller than $G_{0}$. 
Fig. \ref{fig:transport}-(c) shows  $G$ vs $T_{\text{eff}}$ for different values of $T$ and $V$ for the LM, C and MCK fixed points. The grey
curves are obtained by varying $T$ at fixed $V$ for the
lowest value of $V$ considered in our study, {\it i.e.} $V_{\mbox{\tiny min}}=10^{-8}D$. 
The temperature dependence of the linear response conductance is reproduced at all fixed points when the non-linear conductance is taken as a function of $T_{\text{eff}}$. This is true
even for values of $V$ several orders of magnitude larger than $V_{\mbox{\tiny min}}$. 
\newline
{\it In conclusion}, we have addressed the steady-state dynamics near unconventional quantum criticality. We found
that in the scaling regime of all the fixed points considered, all observables studied ($\chi,\phi_s, G$) scale in terms of the same but fixed point specific effective temperature \teff.
The local spin-spin correlation function $\chi$ and the singlet-strength $\phi_s$ assume their equilibrium scaling functions even far from equilibrium when scaled in terms of \teff , {\it i.e.} \teff replacing $T$. A similar result relates the linear and non-linear conductance.
We note that in the (non-interacting) pseudogap resonant level model such behavior is absent~\cite{Zamani}.
It has been shown that the non-equilibrium current noise near quantum criticality in models possessing  gravity duals appears thermal~\cite{Sonner.12,Bhaseen.15}. 
Our results imply that similar results hold for a larger class of quantum critical systems and quantities.
The results reported here may thus help in identifying universality classes of unconventional quantum criticality.
To which extend our results rely on locality needs to be further investigated. 

{\it Acknowledgments.} Helpful discussions with K.~Ingersent, A.~Mitra, and Q.~Si are gratefully acknowledged.
P.\,Ribeiro was supported by the Marie Curie International Reintegration Grant PIRG07-GA-2010-268172. 
S.\,Kirchner acknowledges partial support by the National Natural Science Foundation of China, grant No.11474250.

\newpage 

\begin{widetext}
\begin{center} {\large \bf Steady-state dynamics and effective temperatures\\ of quantum criticality in an open system:\\ \large Supplementary Material }  

\bigskip 

P.~Ribeiro,$^{1,2}$ F.~Zamani,$^{3,4}$ and S.~Kirchner$^{5}$\\
\medskip 
{\it \small
$^1$Russian Quantum Center, Novaya street 100 A, Skolkovo, Moscow area, 143025 Russia\\
$^2$Centro de F\'\i sica das Interac\c c\~oes Fundamentais, Instituto Superior T\'ecnico, Universidade de Lisboa, Av. Rovisco Pais, 1049-001 Lisboa, Portugal\\
$^3$Max Planck Institute for Chemical Physics of Solids, N\"othnitzer Stra\ss{}e 40, 01187 Dresden, Germany\\
$^4$Max Planck Institute for the Physics of Complex Systems, N\"othnitzer Stra\ss{}e 38, 01187 Dresden, Germany\\
$^5$Center for Correlated Matter, Zhejiang University, Hanghzou, Zhejiang 310058, China
}
\end{center}

\setcounter{figure}{0}   \renewcommand{\thefigure}{S\arabic{figure}}
\setcounter{equation}{0} \renewcommand{\theequation}{S.\arabic{equation}}
\setcounter{section}{0} \renewcommand{\thesection}{S.\Roman{section}}
\renewcommand{\thesubsection}{S.\Roman{section}.\Alph{subsection}}
\makeatletter
\renewcommand*{\p@subsection}{}  
\makeatother
\renewcommand{\thesubsubsection}{S.\Roman{section}.\Alph{subsection}-\arabic{subsubsection}}
\makeatletter
\renewcommand*{\p@subsubsection}{}  
\makeatother
\renewcommand*{\citenumfont}[1]{S#1}
\renewcommand*{\bibnumfmt}[1]{[S#1]}
\newcommand{\citesup}[1]{\citetext{S\citealp{#1}}}

\section{Generating functional on the Keldysh contour}

The generating functional on the Keldysh contour
can be written as 
\begin{eqnarray}
Z\left[\xi\right] & = & \int Dc\int Df\int D\lambda e^{i\left(c^{\dagger}g_{c}^{-1}c+f^{\dagger}g_{f}^{-1}f\right)}e^{iqN\int_{\gamma}dz\lambda_{z}}\nonumber \\
 &  & \times e^{+i\frac{1}{N}\int_{\gamma}dz\,\sum_{\alpha}J_{ll'z}\left(\sum_{\sigma}f_{\sigma,z}^{\dagger}c_{0,\alpha\sigma l'z}\right)\left(\sum_{\sigma'}c_{0\alpha\sigma'lz}^{\dagger}f_{\sigma'z}\right)}\nonumber \\
 &  & \times e^{c^{\dagger}\xi_{c}+\xi_{c}^{\dagger}c+f^{\dagger}\xi_{f}+\xi_{f}^{\dagger}f}\label{eq:Z_1}
\end{eqnarray}
where $\xi_{c}$ and $\xi_{f}$ act as sources 
to the fermionic $c$ and $f$ fields and $\lambda$ is a scalar Lagrange multiplier
enforcing the constraint $\sum_{\sigma}f_{\sigma}^{\dagger}f_{\sigma}=qN$. $\int_{\gamma}dz$ is the integral over the Keldysh contour $\gamma=\gamma_{\rightarrow}+\gamma_{\leftarrow}$ with its forward ($\gamma_{\rightarrow}$) and backward  ($\gamma_{\leftarrow}$) branches.

Here, the inverse bare propagators are 
\begin{eqnarray}
g_{f}^{-1} & = & \left(i\pd_{z}-\lambda_{z}\right), \\
g_{c}^{-1} & = & \left(i\pd_{z}-\varepsilon_{pl}\right).
\end{eqnarray}

In analogy to the equilibrium procedure --albeit performed on the Matsubara contour-- 
one can introduce a Hubbard-Stratonovich decoupling field $  B_{\alpha lz} $
conjugated to $\sum_{\sigma'}c_{0\alpha\sigma'lz}^{\dagger}f_{\sigma'z}$,
to decouple the quartic fermionic term in Eq.(\ref{eq:Z_1}). 
Thus,
\begin{eqnarray}
Z\left[\xi\right] & = & \int Df\int D\lambda\int DBe^{i\left(f^{\dagger}g_{f}^{-1}f\right)}e^{i\int dz\lambda_{z}Q}e^{i\left(B_{\alpha}^{\dagger}g_{B}^{-1}B_{\alpha}\right)}\nonumber \\
 &  & \times e^{-i\frac{1}{N}\int dz\int dz'\sum_{\sigma\alpha l}B_{\alpha lz}B_{\alpha lz'}^{\dagger} \tilde{g}_{c,l}(z,z') f_{\sigma,z}^{\dagger}f_{\sigma z'}}\nonumber \\
 &  & \times e^{\left[-\frac{1}{\sqrt{N}}\int_{\gamma}dz'\sum_{\alpha'\sigma'}\left( \xi_{c}^{\dagger}.g_{c} \right)_{0\alpha'\sigma'l'z} B_{\alpha l'z'}^{\dagger}f_{\sigma'z'}-\frac{1}{\sqrt{N}}\int_{\gamma}dz\sum_{\sigma\alpha}f_{\sigma,z}^{\dagger}B_{\alpha lz} \left( g_{c}.\xi_{c}\right)_{0\alpha\sigma lz}  \right]}\nonumber \\
 &  & \times e^{i\xi_{c}^{\dagger}g_{c}\xi_{c}}e^{f^{\dagger}\xi_{f}+\xi_{f}^{\dagger}f}
\end{eqnarray}
with 
\begin{eqnarray}
g_{B; ll'}^{-1} & = & -\left[\tilde{J}^{-1}\right]_{ll'}, 
\end{eqnarray}
where $\left[ \tilde{J} \right]_{ll} = J_{ll'}$, is the bare inverse
propagator of the $B$ field.

Finally, with the help of the complex-valued dynamic Hubbard-Stratonovich fields $W_{zz';l}$ one obtains

\begin{eqnarray}
Z\left[\xi\right] & = & \int D\lambda\int DW\, e^{N\tr\ln\left[-iG_{f}^{-1}\right]-M\tr\ln\left[-i\left(G_{B}^{-1}+V_{\xi_{c}}^{\dagger}G_{f}V_{\xi_{c}}\right)\right]}\nonumber \\
 &  & e^{iN\tr\left[W^{\dagger}*\left[\tilde{g}_{c}\right]^{-1}*W\right]+i\int dz\lambda_{z}Q}\nonumber \\
 &  & e^{-i\xi^\dagger_{f} G_{f}V_{\xi_{c}}\left[G_{B}^{-1}+V_{\xi_{c}}^{\dagger}G_{f}V_{\xi_{c}}\right]^{-1}V_{\xi_{c}}^{\dagger}G_{f}\xi_{f}+i\xi^\dagger_{c}g_{c}\xi_{c}+i\xi^\dagger_{f}G_{f} \xi_{f} }\label{eq:Z:_2}
\end{eqnarray}
with 
\begin{eqnarray}
G_{f}^{-1}(z,z') & = & g_{f}^{-1}(z,z')-W_{zz';l}\\
G_{B}^{-1}(z,z') & = & g_{b}^{-1}(z,z')- \bar{W}_{z'z;l}\\
\tr\left[W^{\dagger}*\left[\tilde{g}_{c}\right]^{-1}*W\right] & = & \sum_{l}\int dz\int dz' \frac{\bar{W}_{zz';l}W_{zz';l}}{\tilde{g}_{c,l}(z,z')}
\end{eqnarray}
and $V_{\xi_{c}}^{\dagger}$ and $V_{\xi_{c}}$ are source-dependent terms. 
Eq.(\ref{eq:Z:_2}) is used to derive all  correlators by taking derivatives with respect to the source fields. 

\section{Dynamical large-N self-consistency equations on the Keldysh contour}

In this section we set the sources to zero and compute the saddle-point
equations with respect to the bosonic fields $W$ and $\lambda$. 
The generating functional in the absence of sources is 
\begin{eqnarray}
Z\left[\xi=0\right] & = & \int D\lambda\int DW\, e^{iN\, S\left[W,\lambda\right]}
\end{eqnarray}
with 
\begin{eqnarray}
S\left[W,\lambda\right] & = & q\,\int dz\lambda_{z}+\tr\left[W^{\dagger}*\left[\tilde{g}_{c}\right]^{-1}*W\right]-i\frac{1}{N}\tr\ln\left[-iG_{f}^{-1}\right]+i\kappa\frac{1}{M}\tr\ln\left[-iG_{B}^{-1}\right].
\end{eqnarray}

The saddle point equations are obtained by putting the linear variation of $S\left[W,\lambda\right]$ with respect to
  $W$ and $\lambda$ to zero:
\begin{eqnarray}
\dfrac{\delta}{\delta W_{zz',l}} S\left[W,\lambda\right] & = & \bar{W}_{zz',l}\left[\tilde{g}_{c}(z,z') \right]^{-1}+i\frac{1}{N}\sum_{\sigma} G_{f}(z',z) = 0,\\
\dfrac{\delta}{\delta \bar{W}_{zz',l}}S\left[W,\lambda\right] & = & \left[ \tilde{g}_{c,l}(z,z') \right]^{-1}W_{zz',l}-i\kappa\frac{1}{M}\sum_{\alpha} G_{B:ll}(z,z') =0,\\
\dfrac{\delta}{\delta \lambda_{z}}S\left[W,\lambda\right] & = & q+i\frac{1}{N}\sum_{\sigma}G_{f}(z^-,z) = 0.
\end{eqnarray}
These equations become {\it exact} in the large-N limit. These equations are equivalent to
\begin{eqnarray*}
\hat{G}_{f}^{-1} & = & \hat{g}_{f}^{-1}-\Sigma_{f}\\
\hat{G}_{B}^{-1} & = & \hat{g}_{B}^{-1}-\Sigma_{B}\\
q & = & -i\hat{G}_{f} (z^-,z)
\end{eqnarray*}
with
\begin{eqnarray}
\Sigma_{B}(z,z') & = & \left(\begin{array}{cc}
\bar{W}_{z'z,L} & 0\\
0 & \bar{W}_{z'z,R}
\end{array}\right)=-i\left(\begin{array}{cc}
\tilde{g}_{c,L}(z',z)  & 0\\
0 & \tilde{g}_{c,R}(z',z)
\end{array}\right) \hat{G}_{f}(z,z')\label{eq:SC_GEN_1}\\
\Sigma_{f}(z,z') & = & \delta_{\sigma\sigma'}\sum_{l}W_{zz',l}=
\delta_{\sigma\sigma'}i\kappa\sum_{l} \tilde{g}_{c,l}(z,z') \hat{G}_{B; ll}(z,z').
\label{eq:SC_GEN_2}
\end{eqnarray}
Note that $\lambda_{z}$ evaluated at the saddle-point
is time independent, {\it i.e.} $\lambda_{t}  =  \lambda$.

\section{Singular exchange coupling matrix $J$}
So far, the treatment has been  general and no particular form of the Kondo exchange coupling matrix has been assumed.
For the physically most relevant case where the 
Kondo Hamiltonian is derived from an Anderson-type model
through a Schrieffer-Wolff transformation, the exchange matrix $J_{ll'}$ ($l,l'=L,R$) 
takes the from
\begin{eqnarray*}
J & = & \left(\begin{array}{cc}
J_{L} & \sqrt{J_{L}J_{R}}\\
\sqrt{J_{L}J_{R}} & J_{R}
\end{array}\right).
\end{eqnarray*}
Thus, the exchange coupling matrix is singular, $\det(J)=0$. In this case, where one of the eigenvalues of $J$ vanishes, we can write
\begin{eqnarray*}
\tilde{J} &=& \ket{u_{+}}\left(J_{L}+J_{R}\right)\bra{u_{+}}
\end{eqnarray*}
with 
\begin{eqnarray*}
\ket{u_{-}} & = & -\sqrt{\frac{J_{R}}{J_{L}+J_{R}}}\ket L+\sqrt{\frac{J_{L}}{J_{L}+J_{R}}}\ket R\\
\ket{u_{+}} & = & \sqrt{\frac{J_{L}}{J_{L}+J_{R}}}\ket L+\sqrt{\frac{J_{R}}{J_{L}+J_{R}}}\ket R.
\end{eqnarray*}
As the exchange matrix is singular, the component $u_{-}$ of the $B$
field has to vanish and thus 
\begin{eqnarray*}
\hat{G}_{B} & = & \ket{u_{+}}\hat{G}_{B+}\bra{u_{+}}.
\end{eqnarray*}
In this case the self-consistent equations simplify to 
\begin{eqnarray}
\Sigma_{B+}(z,z')  & = & -i\tilde{g}_{c,+}(z',z)  \hat{G}_{f}(z,z') ,\label{eq:SC_J_1}\\
\Sigma_{f}(z,z') & = & i\kappa \tilde{g}_{c,+}(z,z') \hat{G}_{B,+}(z,z'),\label{eq:SC_J_2}
\end{eqnarray}
with 
\begin{eqnarray*}
\hat{G}_{B+}^{-1} & = & \hat{g}_{B+}^{-1}-\Sigma_{B},\\
\hat{g}_{B+}^{-1}  & = &  \frac{-1}{J_{L}+J_{R}},\\
\tilde{g}_{c,+} & = & \frac{J_{L}\tilde{g}_{c,L}+J_{R}\tilde{g}_{c,R}}{J_{L}+J_{R}}.
\end{eqnarray*}
Using Langreth's rules,
we obtain 
\begin{eqnarray}
\Sigma_{B}^{>,<}\left(t,t'\right) & = & -i\tilde{g}_{c}^{<,>}\left(t',t\right)\hat{G}_{f}^{>,<}\left(t,t'\right),\label{eq:Self_En_Real_time_1}\\
\Sigma_{f}^{>,<}\left(t,t'\right) & = & i\kappa\tilde{g}_{c}^{>,<}\left(t,t'\right)\hat{G}_{B}^{>,<}\left(t,t'\right),\label{eq:Self_En_Real_time_2}\\
q & = & -i\hat{G}_{f}^{<}\left(0\right).\label{eq:Self_En_Real_time_3}
\end{eqnarray}

\section{Non-equilibirum steady-state (NESS) description}
The steady-state condition implies that the system is time translationally invariant so that
$G^{R,A,K}\left(t,t'\right)=G^{R,A,K}\left(t-t'\right)$.
Therefore, it is advantageous to solve the self-consistent equations 
in the frequency domain.
The conventions of the Fourier transform used by us are  
\begin{eqnarray*}
A\left(t\right) & = & \int\frac{d\omega}{2\pi}A\left(\omega\right)e^{-i\omega t},\\
A\left(\omega\right) & = & \int dt\, A\left(t\right)e^{i\omega t}.
\end{eqnarray*}

Eq.(\ref{eq:Self_En_Real_time_1}-\ref{eq:Self_En_Real_time_3})
take the form
\begin{eqnarray}
\Sigma_{B}^{>,<}\left(\omega\right) & = & -i\int\frac{d\nu}{2\pi}\tilde{g}_{c}^{<,>}\left(\nu-\omega\right)G_{f}^{>,<}\left(\nu\right),\label{eq:Self_En_Real_time_1-1}\\
\Sigma_{f}^{>,<}\left(\omega\right) & = & i\kappa\int\frac{d\nu}{2\pi}\tilde{g}_{c}^{>,<}\left(\omega-\nu\right)G_{B}^{>,<}\left(\nu\right),\label{eq:Self_En_Real_time_2-1}\\
q & = & -i\int\frac{d\omega}{2\pi}G_{f}^{<}\left(\omega\right).\label{eq:Self_En_Real_time_3-1}
\end{eqnarray}

The reservoirs are in equilibrium and are thus characterized by their respective chemical potentials $\mu_L$ and $\mu_R$ and
their respective temperatures $T_L=T_R=T$.
We introduce the following
reservoir quantities 
\begin{eqnarray}
\rho_{c,l}^{\pm}\left(\omega\right) & = & -\frac{1}{2\pi i}\left[\tilde{g}_{c,l}^{>}\left(\omega\right)\pm\tilde{g}_{c,l}^{<}\left(\omega\right)\right],\\
\rho_{c,l}^{H}\left(\omega\right) & = & -\frac{1}{\pi}P\int d\nu\frac{\rho_{c,l}^{-}\left(\nu\right)}{\omega-\nu},
\end{eqnarray}
where $\rho_{c,l}^{-}\left(\omega\right)=\frac{1}{L}\sum_{p}\delta\left(\omega-\varepsilon_{pl}\right)$
is the normalized ($\int d\omega\rho_{c,l}^{-}\left(\omega\right)=1$)
local density of states of reservoir $l$, $\rho_{c,l}^{H}\left(\omega\right)$
is its Hilbert transformed and $\rho_{c,l}^{+}\left(\omega\right)$
is proportional to the Keldysh component of the Green's function.
Since the reservoirs are taken to be in equilibrium, the fluctuation
dissipation theorem can be applied and it is found that 
\begin{eqnarray}
\rho_{c,l}^{+}\left(\omega\right) & = & \text{f}_{l}\left(\omega\right)\rho_{c,l}^{-}\left(\omega\right),
\end{eqnarray}
with 
\begin{eqnarray}
\text{f}_{l}\left(\omega\right) & = & \left[1-2n_{f_{l}}\left(\omega-\mu_{l}\right)\right]=\tanh\left[\frac{\beta_{l}}{2}\left(\omega-\mu_{l}\right)\right].
\end{eqnarray}
Here, $n_{f_{l}}\left(\omega-\mu_{l}\right)$ is the Fermi-function, and $\beta_{l}$
and $\mu_{l}$ are the inverse temperature and the chemical potential
of reservoir $l$. The lead's Green's functions can thus be written
in the form
\begin{eqnarray}
\tilde{g}_{c}^{R,A}\left(\omega\right) & = & -\pi\left[\rho_{c}^{H}\left(\omega\right)\pm i\rho_{c}^{-}\left(\omega\right)\right],\\
\tilde{g}_{c}^{K}\left(\omega\right) & = & -2\pi i\rho_{c}^{+}\left(\omega\right),
\end{eqnarray}
with 
\begin{eqnarray}
\rho_{c}^{-}\left(\omega\right) & = & \frac{J_{L}\rho_{c,L}^{-}\left(\omega\right)+J_{R}\rho_{c,R}^{-}\left(\omega\right)}{J_{L}+J_{R}},\\
\rho_{c}^{+}\left(\omega\right) & = & \frac{J_{L}\text{f}_{L}\left(\omega\right)\rho_{c,L}^{-}\left(\omega\right)+J_{R}\text{f}_{R}\left(\omega\right)\rho_{c,R}^{-}\left(\omega\right)}{J_{L}+J_{R}}.
\end{eqnarray}

\subsubsection{Self-consistent equations for the steady-state}

With the definitions of the previous sections, Dyson's equation translates
to 
\begin{eqnarray*}
\rho_{f}^{\pm}\left(\omega\right) & = & \sigma_{f}^{\pm}\left(\omega\right)\left\{ \left[\omega-\tilde{\lambda}+\pi\sigma_{f}^{H}\left(\omega\right)\right]^{2}+\left[\pi\sigma_{f}^{-}\left(\omega\right)\right]^{2}\right\} ^{-1},\\
\rho_{B}^{\pm}\left(\omega\right) & = & \sigma_{B}^{\pm}\left(\omega\right)\left\{ \left[-\left(J_{L}+J_{R}\right)+\pi\sigma_{B}^{H}\left(\nu\right)\right]^{2}+\left[\pi\sigma_{B}^{-}\left(\omega\right)\right]^{2}\right\} ^{-1},
\end{eqnarray*}
with $\tilde{\lambda}=\lambda_{t}-\frac{\kappa}{2}\left(J_{L}+J_{R}\right)\int d\omega\rho_{c}^{+}\left(\omega\right)$ being
a renormalized chemical potential, and Eq.(\ref{eq:Self_En_Real_time_1-1}-\ref{eq:Self_En_Real_time_3-1})
translate to
\begin{eqnarray}
\sigma_{B}^{\pm}\left(\omega\right) & = & \mp\frac{1}{2}\int d\nu\left[\rho_{c}^{\pm}\left(\nu-\omega\right)\rho_{f}^{+}\left(\nu\right)-\rho_{c}^{\mp}\left(\nu-\omega\right)\rho_{f}^{-}\left(\nu\right)\right],\label{eq:Self_En_SS_1}\\
\sigma_{f}^{\pm}\left(\omega\right) & = & \kappa\frac{1}{2}\int d\nu\left[\rho_{c}^{\pm}\left(\omega-\nu\right)\rho_{B}^{+}\left(\nu\right)+\rho_{c}^{\mp}\left(\omega-\nu\right)\rho_{B}^{-}\left(\nu\right)\right],\label{eq:Self_En_SS_2}\\
q & = & \frac{1}{2}\left[1-\int d\omega\rho_{f}^{+}\left(\omega\right)\right].
\end{eqnarray}
In the particle-hole symmetric case ($q=1/2$) and for a particle-hole symmetric DOS
of the leads ($\rho_{c}^{-}\left(\omega\right)=\rho_{c}^{-}\left(-\omega\right)$)
the  quantities $\rho_{f,B}^{\pm}$ and $\sigma_{f,B}^{\pm}$ are
real.

\subsubsection{Details of the numerical treatment}

The explicit form of the pseudogap density of states of the 
leads is taken to be 
\begin{eqnarray*}
\rho_{c,l}^{-}\left(\omega\right) & = & \frac{1}{\sqrt{2}\Lambda\Gamma\left(\frac{r+1}{2}\right)}\left|\frac{\omega}{\sqrt{2}\Lambda}\right|^{r}e^{-\frac{\omega^{2}}{2\Lambda^{2}}},
\end{eqnarray*}
with $l=R,L$ and $\Lambda=1$ specifies the soft  high-energy cutoff. 
The self-consistent equations were solved iteratively on a logarithmically
discretized grid with $350$ points ranging from $-10\Lambda$ to
$10\Lambda$. The criterium for convergence of the selfconsistency loop was that the relative
difference of two consecutive iterations was less than \textbf{$10^{-6}$}.
The results were benchmarked by the conditions that the fluctuation dissipation ratios of the Green's functions have to accurately
reproduce the equilibrium fluctuation dissipation relations  demanded by the fluctuation-dissipation theorem. 
For all the fixed points we studied a range of temperatures $T/D = 10^{-1}, 10^{-1.5}, 10^{-2}, \ldots ,10^{-8} $ and a range of voltages $T/D = 10^{-2}, 10^{-1.5}, 10^{-2}, \ldots ,10^{-8} $. However convergence of the numerical solution of the self-consistent equations was not always achieved for all combinations of parameters. 

\section{Observables}

\subsubsection{Cross 4-point function}

In order to compute the currents and the Kondo singlet strength we will
need to evaluate the connected 4-point function $\av{T_{\gamma}c_{p_{2}\alpha_{2}\sigma_{2}l_{2}}^{\dagger}c_{p_{1}\alpha_{1}\sigma_{1}l_{1}}f_{s_{2}}^{\dagger}f_{s_{1}}}_{C}$.
Here, $C$ denotes the connected part of a correlation function and $T_{\gamma}$ is the time-ordering operator on the Keldysh contour.
Using the procedure outlined above, one obtains
\begin{eqnarray*}
\av{T_{\gamma}f_{s_{1}}\left(t_{1}\right)f_{s_{2}}^{\dagger}\left(t_{2}\right)c_{p_{1}\alpha_{1}\sigma_{1}l_{1}}\left(t_{3}\right)c_{p_{2}\alpha_{2}\sigma_{2}l_{2}}^{\dagger}\left(t{}_{4}\right)}_{C} & = & i\frac{1}{N}\frac{1}{n_{c}}\delta_{s_{1}\sigma_{2}}\delta_{s_{2}\sigma_{1}}\delta_{\alpha_{1}\alpha_{2}}F_{p_{1}l_{1};p_{2}l_{2}}\left(t_{1},t_{2},t_{3},t_{4}\right)
\end{eqnarray*}
and 
\begin{eqnarray*}
F_{p_{1}l_{1};p_{2}l_{2}}\left(t_{1},t_{2},t_{3},t_{4}\right) & = & \frac{\sqrt{J_{l_{2}}J_{l_{1}}}}{\sqrt{\left(J_{L}+J_{R}\right)\left(J_{L}+J_{R}\right)}}\\
 &  & \times\int dz'\int dz\ G_{f}\left(t_{1},z\right)g_{c;p_{2}l_{2}}\left(z,t{}_{4}\right)G_{B}\left(z,z'\right)g_{c;p_{1}l_{1}}\left(t{}_{3},z'\right)G_{f}\left(z',t_{2}\right)
\end{eqnarray*}
with $g_{c;p_{1}l_{1}}\left(t,t'\right)=\bra{tp_{1}l_{1}}g_{c}\ket{t'p_{1}l_{1}}$.
For equal times we have 
\begin{eqnarray*}
\av{c_{p_{2}\alpha_{2}\sigma_{2}l_{2}}^{\dagger}\left(t\right)c_{p_{1}\alpha_{1}\sigma_{1}l_{1}}\left(t\right)f_{s_{2}}^{\dagger}\left(t\right)f_{s_{1}}\left(t\right)}_{C} & = & i\frac{1}{N}\frac{1}{n_{c}}\delta_{s_{1}\sigma_{2}}\delta_{s_{2}\sigma_{1}}\delta_{\alpha_{1}\alpha_{2}}F_{p_{1}l_{1};p_{2}l_{2}}\left(t\right),
\end{eqnarray*}
where the time-ordering for the equal-time limit is defined through $F_{p_{1}l_{1};p_{2}l_{2}}\left(t\right)=\left.\lim_{t_{1,2,3,4}\to t}F_{p_{1}l_{1};p_{2}l_{2}}\left(t_{1},t_{2},t_{3},t_{4}\right)\right|_{t_{1}>t_{2}>t_{3}>t_{4}}$.
$F_{p_{1}l_{1};p_{2}l_{2}}\left(t\right)$ can be explicitly evaluated
using Langreth rules and making use of the  fact that we describe a steady-state. This procedure
is straightforward but involved and yields 
\begin{eqnarray*}
F_{p_{1}l_{1};p_{2}l_{2}}\left(t\right) & = & 4i\pi^{5}\left[\mathcal{I}_{l_{1}p_{1},l_{2}p_{2}}^{\left(1\right)}+\mathcal{\mathcal{I}}_{l_{1}p_{1},l_{2}p_{2}}^{\left(2\right)}\right],
\end{eqnarray*}
with 
\begin{eqnarray*}
\mathcal{I}_{p_{1},p_{2}}^{\left(1\right)} & = & \frac{1}{8}\frac{\sqrt{J_{l_{2}}J_{l_{1}}}}{\left(J_{L}+J_{R}\right)}\int\frac{d\omega}{2\pi}\left\{ \left[-i\mathcal{H}\left[A_{l_{2}}^{-++-}\right]\left(\omega\right)+A_{l_{2}}^{-++-}\left(\omega\right)\right]\left[2\rho_{B}^{+}(\omega)\right]\left[i\mathcal{H}\left[A_{l_{1}}^{-++-}\right]\left(\omega\right)+A_{l_{1}}^{-++-}\left(\omega\right)\right]\right.\\
 &  & +\left[-i\mathcal{H}\left[A_{l_{2}}^{-++-}\right]\left(\omega\right)+A_{l_{2}}^{-++-}\left(\omega\right)\right]\left[-i\rho_{B}^{H}\left(\omega\right)+\rho_{B}^{-}\left(\omega\right)\right]\left[i\mathcal{H}\left[A_{l_{1}}^{--++}\right]\left(\omega\right)+A_{l_{1}}^{--++}\left(\omega\right)\right]\\
 &  & \left.+\left[-i\mathcal{H}\left[A_{l_{2}}^{--++}\right]\left(\omega\right)+A_{l_{2}}^{--++}\left(\omega\right)\right]\left[i\rho_{B}^{H}\left(\omega\right)+\rho_{B}^{-}\left(\omega\right)\right]\left[i\mathcal{H}\left[A_{l_{1}}^{-++-}\right]\left(\omega\right)+A_{l_{1}}^{-++-}\left(\omega\right)\right]\right\} \\
\mathcal{I}_{p_{1},p_{2}}^{\left(2\right)} & = & \frac{1}{2}\sqrt{J_{l_{2}}J_{l_{1}}}\frac{1}{2\pi i}\int\frac{d\omega}{2\pi}\left\{ \left[-i\mathcal{H}\left[A_{l_{2}}^{-++-}\right]\left(\omega\right)+A_{l_{2}}^{-++-}\left(\omega\right)\right]A_{l_{1}}^{--++}\left(\omega\right)\right.\\
 &  & \left.-\left[-i\mathcal{H}\left[A_{l_{2}}^{--++}\right]\left(\omega\right)+A_{l_{2}}^{--++}
 \left(\omega\right)\right]A_{l_{1}}^{-++-}\left(\omega\right)\right\},
\end{eqnarray*}
where we defined 
\begin{eqnarray*}
A_{l}^{\Sigma}\left(\omega\right) & = & \int\frac{d\nu}{2\pi}\left[\rho_{f}^{\Sigma(1)}\left(\nu\right)
\rho_{c,p_{l}}^{\Sigma(2)}\left(\nu-\omega\right)-\rho_{f}^{\Sigma(3)}\left(\nu\right)
\rho_{c,p_{l}}^{\Sigma(4)}\left(\nu-\omega\right)\right],\\
\mathcal{H}\left[A\right]\left(\omega\right) & = & -\frac{1}{\pi}P\int d\nu\frac{A\left(\nu\right)}{\omega-\nu}.
\end{eqnarray*}

\subsubsection{Currents}

The currents of particles and energy through the system are obtained from the change in particle number and energy of {\it e.g.}
the left lead through a
 continuity equation for the conserved charge (particle number or energy),
\begin{eqnarray*}
\mathcal{J}_{b} & = & -\partial_{t}\av{\mathcal{Q}_{b}\left(t\right)}=-i\av{\left[H\left(t\right),\mathcal{Q}_{b}\left(t\right)\right]}\\
\mathcal{J}_{\text{E}} & \to & \mathcal{Q}_{\text{E}}=H_{L}=\sum_{p,\alpha\sigma}\varepsilon_{pL}c_{p\alpha\sigma L}^{\dagger}c_{p\alpha\sigma L}\\
\mathcal{J}_{\text{P}} & \to & \mathcal{Q}_{\text{P}}=N_{L}=\sum_{p,\alpha\sigma}c_{p\alpha\sigma L}^{\dagger}c_{p\alpha\sigma L}
\end{eqnarray*}
Using the identity $\left[c_{\alpha}^{\dagger}\mathit{c}_{\beta},c_{\gamma}^{\dagger}\mathit{c}_{\delta}\right]=\delta_{\beta,\gamma}c_{\alpha}^{\dagger}\mathit{c}_{\delta}-\delta_{\alpha,\delta}c_{\gamma}^{\dagger}\mathit{c}_{\beta}$
and the fact that the Hamiltonian can be decomposed as $H=H_{L}+H_{R}+H_{J}$
with 
\begin{eqnarray*}
H_{J} & = & \frac{1}{N}\sum_{ll'}J_{ll'}\sum_{\sigma\sigma'}\sum_{\alpha}\left(f_{\sigma}^{\dagger}\, f_{\sigma'}-q\delta_{\sigma\sigma'}\right)c_{0,\alpha\sigma'l'}^{\dagger}c_{0\alpha\sigma l},
\end{eqnarray*}

one obtains 
\begin{eqnarray*}
\mathcal{J}_{\text{P}}\left(t\right)/M & = & 2\sqrt{J_{L,t}J_{R,t}}\re\left[\frac{1}{n_{c}^{2}}\sum_{pp'}F_{Rp',Lp}\left(t\right)\right],\\
\mathcal{J}_{\text{E}}\left(t\right)/M & = & 2\sqrt{J_{L,t}J_{R,t}}\re\left[\frac{1}{n_{c}^{2}}\sum_{pp'}\varepsilon_{pL}F_{Rp',Lp}\left(t\right)\right].
\end{eqnarray*}

\subsubsection{Susceptibility}

On the Keldysh contour the impurity spin susceptibility is defined
by 
\begin{eqnarray*}
\chi\left(z,z'\right) & = & -i\frac{1}{N}\sum_{a}\av{T_{\gamma}S^{a}\left(z\right)S^{a}\left(z'\right)},
\end{eqnarray*}
where $T_{\gamma}$ is the time-ordering operator on the Keldysh contour.

For a steady state, we obtain 
\begin{eqnarray*}
\chi^{\pm}\left(\omega\right) & = & -\frac{1}{2}\int d\nu\left[\rho_{f}^{+}\left(\nu-\omega\right)\rho_{f}^{\pm}\left(\nu\right)-\rho_{f}^{-}
\left(\omega-\nu\right)\rho_{f}^{\mp}\left(\nu\right)\right],
\end{eqnarray*}
where $\chi_{f}^{\pm}\left(\omega\right)=-\frac{1}{2\pi i}\left[\chi^{>}\left(\omega\right)\pm\chi^{<}\left(\omega\right)\right]$.

\subsubsection{Kondo singlet strength}

It follows from the Hamiltonian, Eq.~(\ref{eq:Hamiltonian_PGK}), that the Kondo  term contribution to the total energy is given by

\begin{eqnarray*}
E_{K}\left(t\right) & = & \frac{1}{N}\sum_{ll'}\sum_{c}J_{ll'}\av{\mathbf{S}\left(t\right).\mathbf{s}_{c,ll'}\left(t\right)}\\
 & = & \kappa\left(\frac{N^{2}-1}{N}\right)\left\{ i\sum_{l_{1}l_{2}}J_{l_{1}l_{2}}\left[\frac{1}{n_{c}^{2}}\sum_{p_{1}p_{2}}F_{p_{1}l_{1};p_{2}l_{2}}\left(t\right)\right]\right\} \\
 & = & -J\kappa\left(\frac{N^{2}-1}{N}\right)\phi_{s}\left(t\right).
\end{eqnarray*}
This  expression can be greatly simplified using the definition
of $F_{p_{1}l_{1};p_{2}l_{2}}\left(t\right)$, see previous section. This then yields for the Kondo singlet strength 
\begin{eqnarray*}
\phi_{s} & = & \pi/J\, w^{H}\left(0\right),
\end{eqnarray*}
with
\begin{eqnarray*}
w^{-}\left(\omega\right) & = & \frac{1}{2}\int d\nu\left[\sigma_{B}^{+}\left(\nu-\omega\right)\rho_{B}^{-}\left(\nu\right)-\sigma_{B}^{-}
\left(\nu-\omega\right)\rho_{B}^{+}\left(\nu\right)\right].
\end{eqnarray*}

\section{Additional numerical results - Other values of $r$ and $\kappa$}

In this section we provide further numerical support for our conclusions. 
Figure \ref{fig:effectiveT2} shows our results for the parameter set $(r,\kappa) = (0.15,0.16)$ which is different from the one the results in the paper are based on.

%
\begin{figure}
\centering{}\includegraphics[width=0.9\linewidth]{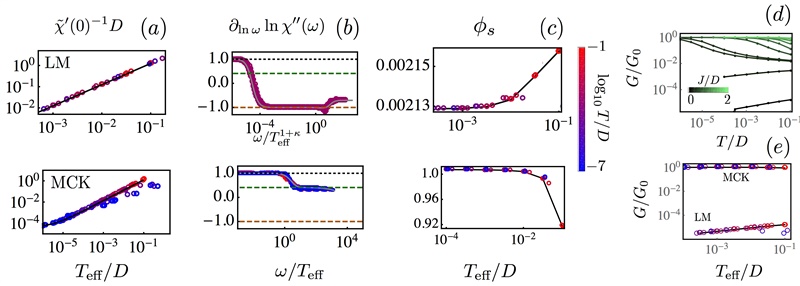}\protect\caption{\label{fig:effectiveT2}
Scaling of different observables with $T_{\text{Jeff}}$ for the different fixed points(the parameters used here differ from those of Figure 3 and 4 of the paper):
(a) Inverse static susceptibility $\chi'\left(0\right)^{-1}$vs
$T_{\text{eff}}$ ; (b) $\protect\pd_{\ln\omega}\ln\chi''\left(\omega\right)$
vs $\omega/T_{\text{eff}}$; (c) singlet strength $\phi_{s}$ vs $T_{\text{eff}}$
For each fixed point, the equilibrium scaling form (black dashed lines)
is compared with the same quantity under non-equilibrium conditions
where $T$ is substituted by $T_{\text{eff}}$.
(d) Conductance $G$ as a function of temperature computed for the lowest non-zero value of $V$ for
several values of $J$ (see color coding). 
(e)  $G=\mathcal{J}_{\text{P}}/V$ vs $T_{\text{eff}}$. for the different fixed points. The equilibrium form is depicted by the black dashed lines.
$G_0$ is defined as the zero-temperature limit of $G$ in the MCK regime.
}
\end{figure}

\newpage\end{widetext}


\begin{thebibliography}{56}
\expandafter\ifx\csname natexlab\endcsname\relax\def\natexlab#1{#1}\fi
\expandafter\ifx\csname bibnamefont\endcsname\relax
  \def\bibnamefont#1{#1}\fi
\expandafter\ifx\csname bibfnamefont\endcsname\relax
  \def\bibfnamefont#1{#1}\fi
\expandafter\ifx\csname citenamefont\endcsname\relax
  \def\citenamefont#1{#1}\fi
\expandafter\ifx\csname url\endcsname\relax
  \def\url#1{\texttt{#1}}\fi
\expandafter\ifx\csname urlprefix\endcsname\relax\def\urlprefix{URL }\fi
\providecommand{\bibinfo}[2]{#2}
\providecommand{\eprint}[2][]{\url{#2}}

\bibitem[{\citenamefont{Eckstein et~al.}(2010)\citenamefont{Eckstein, Hackl,
  Kehrein, Kollar, Moeckel, Werner, and Wolf}}]{Eckstein2010}
\bibinfo{author}{\bibfnamefont{M.}~\bibnamefont{Eckstein}},
  \bibinfo{author}{\bibfnamefont{A.}~\bibnamefont{Hackl}},
  \bibinfo{author}{\bibfnamefont{S.}~\bibnamefont{Kehrein}},
  \bibinfo{author}{\bibfnamefont{M.}~\bibnamefont{Kollar}},
  \bibinfo{author}{\bibfnamefont{M.}~\bibnamefont{Moeckel}},
  \bibinfo{author}{\bibfnamefont{P.}~\bibnamefont{Werner}}, \bibnamefont{and}
  \bibinfo{author}{\bibfnamefont{F.}~\bibnamefont{Wolf}}, \bibinfo{journal}{The
  European Physical Journal Special Topics} \textbf{\bibinfo{volume}{180}},
  \bibinfo{pages}{217} (\bibinfo{year}{2010}).

\bibitem[{\citenamefont{Arrigoni et~al.}(2013)\citenamefont{Arrigoni, Knap, and
  von~der Linden}}]{Arrigoni2013}
\bibinfo{author}{\bibfnamefont{E.}~\bibnamefont{Arrigoni}},
  \bibinfo{author}{\bibfnamefont{M.}~\bibnamefont{Knap}}, \bibnamefont{and}
  \bibinfo{author}{\bibfnamefont{W.}~\bibnamefont{von~der Linden}},
  \bibinfo{journal}{Phys.~Rev.~Lett.} \textbf{\bibinfo{volume}{110}},
  \bibinfo{pages}{086403} (\bibinfo{year}{2013}).

\bibitem[{\citenamefont{{Mu\~noz} et~al.}(2013)\citenamefont{{Mu\~noz}, Bolech,
  and Kirchner}}]{Munoz.13}
\bibinfo{author}{\bibfnamefont{E.}~\bibnamefont{{Mu\~noz}}},
  \bibinfo{author}{\bibfnamefont{C.~J.} \bibnamefont{Bolech}},
  \bibnamefont{and} \bibinfo{author}{\bibfnamefont{S.}~\bibnamefont{Kirchner}},
  \bibinfo{journal}{Phys.~Rev.~Lett.} \textbf{\bibinfo{volume}{110}},
  \bibinfo{pages}{016601} (\bibinfo{year}{2013}).

\bibitem[{\citenamefont{Werner et~al.}(2009)\citenamefont{Werner, Oka, and
  Millis}}]{Werner2009}
\bibinfo{author}{\bibfnamefont{P.}~\bibnamefont{Werner}},
  \bibinfo{author}{\bibfnamefont{T.}~\bibnamefont{Oka}}, \bibnamefont{and}
  \bibinfo{author}{\bibfnamefont{A.}~\bibnamefont{Millis}},
  \bibinfo{journal}{Phys.~Rev.~B} \textbf{\bibinfo{volume}{79}},
  \bibinfo{pages}{035320} (\bibinfo{year}{2009}).

\bibitem[{\citenamefont{Gull et~al.}(2011)\citenamefont{Gull, Reichman, and
  Millis}}]{Gull2011}
\bibinfo{author}{\bibfnamefont{E.}~\bibnamefont{Gull}},
  \bibinfo{author}{\bibfnamefont{D.~R.} \bibnamefont{Reichman}},
  \bibnamefont{and} \bibinfo{author}{\bibfnamefont{A.~J.}
  \bibnamefont{Millis}}, \bibinfo{journal}{Phys.~ Rev.~B}
  \textbf{\bibinfo{volume}{84}}, \bibinfo{pages}{085134}
  (\bibinfo{year}{2011}).

\bibitem[{\citenamefont{Cohen et~al.}(2013)\citenamefont{Cohen, Gull, Reichman,
  Millis, and Rabani}}]{Cohen2013}
\bibinfo{author}{\bibfnamefont{G.}~\bibnamefont{Cohen}},
  \bibinfo{author}{\bibfnamefont{E.}~\bibnamefont{Gull}},
  \bibinfo{author}{\bibfnamefont{D.~R.} \bibnamefont{Reichman}},
  \bibinfo{author}{\bibfnamefont{A.~J.} \bibnamefont{Millis}},
  \bibnamefont{and} \bibinfo{author}{\bibfnamefont{E.}~\bibnamefont{Rabani}},
  \bibinfo{journal}{Phys.~Rev.~B} \textbf{\bibinfo{volume}{87}},
  \bibinfo{pages}{195108} (\bibinfo{year}{2013}).

\bibitem[{\citenamefont{Aoki et~al.}(2014)\citenamefont{Aoki, Tsuji, Eckstein,
  Kollar, Oka, and Werner}}]{Aoki2014}
\bibinfo{author}{\bibfnamefont{H.}~\bibnamefont{Aoki}},
  \bibinfo{author}{\bibfnamefont{N.}~\bibnamefont{Tsuji}},
  \bibinfo{author}{\bibfnamefont{M.}~\bibnamefont{Eckstein}},
  \bibinfo{author}{\bibfnamefont{M.}~\bibnamefont{Kollar}},
  \bibinfo{author}{\bibfnamefont{T.}~\bibnamefont{Oka}}, \bibnamefont{and}
  \bibinfo{author}{\bibfnamefont{P.}~\bibnamefont{Werner}},
  \bibinfo{journal}{Rev.~Mod.~Phys.} \textbf{\bibinfo{volume}{86}},
  \bibinfo{pages}{779} (\bibinfo{year}{2014}).

\bibitem[{\citenamefont{Schir\'{o} and Fabrizio}(2010)}]{Schiro2010}
\bibinfo{author}{\bibfnamefont{M.}~\bibnamefont{Schir\'{o}}} \bibnamefont{and}
  \bibinfo{author}{\bibfnamefont{M.}~\bibnamefont{Fabrizio}},
  \bibinfo{journal}{Phys.~Rev.~Lett.} \textbf{\bibinfo{volume}{105}},
  \bibinfo{pages}{076401} (\bibinfo{year}{2010}).

\bibitem[{\citenamefont{Rosch}(2012)}]{Rosch.12}
\bibinfo{author}{\bibfnamefont{A.}~\bibnamefont{Rosch}},
  \bibinfo{journal}{Eur.~Phys.~J.~B} \textbf{\bibinfo{volume}{85}},
  \bibinfo{pages}{6} (\bibinfo{year}{2012}).

\bibitem[{\citenamefont{Nghiem and Costi}(2014{\natexlab{a}})}]{Nghiem.14}
\bibinfo{author}{\bibfnamefont{H.~T.~M.} \bibnamefont{Nghiem}}
  \bibnamefont{and} \bibinfo{author}{\bibfnamefont{T.~A.} \bibnamefont{Costi}},
  \bibinfo{journal}{Phys.~Rev.~B} \textbf{\bibinfo{volume}{90}},
  \bibinfo{pages}{035129} (\bibinfo{year}{2014}{\natexlab{a}}).

\bibitem[{\citenamefont{Nghiem and Costi}(2014{\natexlab{b}})}]{Nghiem.a.14}
\bibinfo{author}{\bibfnamefont{H.~T.~M.} \bibnamefont{Nghiem}}
  \bibnamefont{and} \bibinfo{author}{\bibfnamefont{T.~A.} \bibnamefont{Costi}},
  \bibinfo{journal}{Phys.~Rev.~B} \textbf{\bibinfo{volume}{89}},
  \bibinfo{pages}{075118} (\bibinfo{year}{2014}{\natexlab{b}}).

\bibitem[{\citenamefont{Mitra et~al.}(2006)\citenamefont{Mitra, Takei, Kim, and
  Millis}}]{Mitra.06}
\bibinfo{author}{\bibfnamefont{A.}~\bibnamefont{Mitra}},
  \bibinfo{author}{\bibfnamefont{S.}~\bibnamefont{Takei}},
  \bibinfo{author}{\bibfnamefont{Y.~B.} \bibnamefont{Kim}}, \bibnamefont{and}
  \bibinfo{author}{\bibfnamefont{A.~J.} \bibnamefont{Millis}},
  \bibinfo{journal}{Phys.~Rev.~Lett.} \textbf{\bibinfo{volume}{97}},
  \bibinfo{pages}{236808} (\bibinfo{year}{2006}).

\bibitem[{\citenamefont{Diehl et~al.}(2008)\citenamefont{Diehl, Micheli,
  Kantian, Kraus, B\"{u}chler, and Zoller}}]{Diehl2008}
\bibinfo{author}{\bibfnamefont{S.}~\bibnamefont{Diehl}},
  \bibinfo{author}{\bibfnamefont{A.}~\bibnamefont{Micheli}},
  \bibinfo{author}{\bibfnamefont{A.}~\bibnamefont{Kantian}},
  \bibinfo{author}{\bibfnamefont{B.}~\bibnamefont{Kraus}},
  \bibinfo{author}{\bibfnamefont{H.~P.} \bibnamefont{B\"{u}chler}},
  \bibnamefont{and} \bibinfo{author}{\bibfnamefont{P.}~\bibnamefont{Zoller}},
  \bibinfo{journal}{Nature Phys.} \textbf{\bibinfo{volume}{4}},
  \bibinfo{pages}{878} (\bibinfo{year}{2008}).

\bibitem[{\citenamefont{Hogan and Green}(2008)}]{Hogan.08}
\bibinfo{author}{\bibfnamefont{P.~M.} \bibnamefont{Hogan}} \bibnamefont{and}
  \bibinfo{author}{\bibfnamefont{A.~G.} \bibnamefont{Green}},
  \bibinfo{journal}{Phys.~Rev.~B} \textbf{\bibinfo{volume}{78}},
  \bibinfo{pages}{195104} (\bibinfo{year}{2008}).

\bibitem[{\citenamefont{Chung et~al.}(2009)\citenamefont{Chung, {Le Hur},
  Vojta, and W\"{o}lfle}}]{Chung2009}
\bibinfo{author}{\bibfnamefont{C.-H.} \bibnamefont{Chung}},
  \bibinfo{author}{\bibfnamefont{K.}~\bibnamefont{{Le Hur}}},
  \bibinfo{author}{\bibfnamefont{M.}~\bibnamefont{Vojta}}, \bibnamefont{and}
  \bibinfo{author}{\bibfnamefont{P.}~\bibnamefont{W\"{o}lfle}},
  \bibinfo{journal}{Phys.~Rev.~Lett.} \textbf{\bibinfo{volume}{102}},
  \bibinfo{pages}{216803} (\bibinfo{year}{2009}).

\bibitem[{\citenamefont{Kirchner and Si}(2009)}]{Kirchner.09}
\bibinfo{author}{\bibfnamefont{S.}~\bibnamefont{Kirchner}} \bibnamefont{and}
  \bibinfo{author}{\bibfnamefont{Q.}~\bibnamefont{Si}},
  \bibinfo{journal}{Phys.~Rev.~Lett.} \textbf{\bibinfo{volume}{103}},
  \bibinfo{pages}{206401} (\bibinfo{year}{2009}).

\bibitem[{\citenamefont{Takei et~al.}(2010)\citenamefont{Takei, Witczak-Krempa,
  and Kim}}]{Takei2010}
\bibinfo{author}{\bibfnamefont{S.}~\bibnamefont{Takei}},
  \bibinfo{author}{\bibfnamefont{W.}~\bibnamefont{Witczak-Krempa}},
  \bibnamefont{and} \bibinfo{author}{\bibfnamefont{Y.~B.} \bibnamefont{Kim}},
  \bibinfo{journal}{Phys.~Rev.~B} \textbf{\bibinfo{volume}{81}},
  \bibinfo{pages}{125430} (\bibinfo{year}{2010}).

\bibitem[{\citenamefont{Ribeiro et~al.}(2013)\citenamefont{Ribeiro, Si, and
  Kirchner}}]{Ribeiro2013b}
\bibinfo{author}{\bibfnamefont{P.}~\bibnamefont{Ribeiro}},
  \bibinfo{author}{\bibfnamefont{Q.}~\bibnamefont{Si}}, \bibnamefont{and}
  \bibinfo{author}{\bibfnamefont{S.}~\bibnamefont{Kirchner}},
  \bibinfo{journal}{Europhys.~Lett.} \textbf{\bibinfo{volume}{102}},
  \bibinfo{pages}{50001} (\bibinfo{year}{2013}).

\bibitem[{\citenamefont{Sieberer et~al.}(2013)\citenamefont{Sieberer, Huber,
  Altman, and Diehl}}]{Sieberer2013}
\bibinfo{author}{\bibfnamefont{L.~M.} \bibnamefont{Sieberer}},
  \bibinfo{author}{\bibfnamefont{S.~D.} \bibnamefont{Huber}},
  \bibinfo{author}{\bibfnamefont{E.}~\bibnamefont{Altman}}, \bibnamefont{and}
  \bibinfo{author}{\bibfnamefont{S.}~\bibnamefont{Diehl}},
  \bibinfo{journal}{Phys.~Rev.~Lett.} \textbf{\bibinfo{volume}{110}},
  \bibinfo{pages}{195301} (\bibinfo{year}{2013}).

\bibitem[{\citenamefont{Hohenberg and Halperin}(1977)}]{Hohenberg.77}
\bibinfo{author}{\bibfnamefont{P.~C.} \bibnamefont{Hohenberg}}
  \bibnamefont{and} \bibinfo{author}{\bibfnamefont{B.~I.}
  \bibnamefont{Halperin}}, \bibinfo{journal}{Rev.~Mod.~Phys.}
  \textbf{\bibinfo{volume}{49}}, \bibinfo{pages}{435} (\bibinfo{year}{1977}).

\bibitem[{\citenamefont{Hohenberg and Shariman}(1989)}]{Hohenberg1989}
\bibinfo{author}{\bibfnamefont{P.}~\bibnamefont{Hohenberg}} \bibnamefont{and}
  \bibinfo{author}{\bibfnamefont{B.~I.} \bibnamefont{Shariman}},
  \bibinfo{journal}{Physica D: Nonlinear Phenomena}
  \textbf{\bibinfo{volume}{37}}, \bibinfo{pages}{109} (\bibinfo{year}{1989}).

\bibitem[{\citenamefont{Cugliandolo et~al.}(1997)\citenamefont{Cugliandolo,
  Kurchan, and Peliti}}]{Cugliandolo.97}
\bibinfo{author}{\bibfnamefont{L.~F.} \bibnamefont{Cugliandolo}},
  \bibinfo{author}{\bibfnamefont{J.}~\bibnamefont{Kurchan}}, \bibnamefont{and}
  \bibinfo{author}{\bibfnamefont{L.}~\bibnamefont{Peliti}},
  \bibinfo{journal}{Phys.~Rev.~E} \textbf{\bibinfo{volume}{55}},
  \bibinfo{pages}{3898} (\bibinfo{year}{1997}).

\bibitem[{\citenamefont{Calabrese and Gambassi}(2004)}]{Calabrese.04}
\bibinfo{author}{\bibfnamefont{P.}~\bibnamefont{Calabrese}} \bibnamefont{and}
  \bibinfo{author}{\bibfnamefont{A.}~\bibnamefont{Gambassi}},
  \bibinfo{journal}{J.~Stat.~Mech.} p. \bibinfo{pages}{P07013}
  (\bibinfo{year}{2004}).

\bibitem[{\citenamefont{Bonart et~al.}(2012)\citenamefont{Bonart, Cugliandolo,
  and Gambassi}}]{Bonart.12}
\bibinfo{author}{\bibfnamefont{J.}~\bibnamefont{Bonart}},
  \bibinfo{author}{\bibfnamefont{L.~F.} \bibnamefont{Cugliandolo}},
  \bibnamefont{and} \bibinfo{author}{\bibfnamefont{A.}~\bibnamefont{Gambassi}},
  \bibinfo{journal}{J.~Stat.~Mech.} \textbf{\bibinfo{volume}{2012}},
  \bibinfo{pages}{P01014} (\bibinfo{year}{2012}).

\bibitem[{\citenamefont{Cugliandolo}(2011)}]{Cugliandolo.11}
\bibinfo{author}{\bibfnamefont{L.~F.} \bibnamefont{Cugliandolo}},
  \bibinfo{journal}{J.~Phys.~A:Math.~Theor.} \textbf{\bibinfo{volume}{44}},
  \bibinfo{pages}{483001} (\bibinfo{year}{2011}).

\bibitem[{\citenamefont{Mitra and Millis}(2005)}]{Mitra2005}
\bibinfo{author}{\bibfnamefont{A.}~\bibnamefont{Mitra}} \bibnamefont{and}
  \bibinfo{author}{\bibfnamefont{A.~J.} \bibnamefont{Millis}},
  \bibinfo{journal}{Phys.~Rev.~B} \textbf{\bibinfo{volume}{72}},
  \bibinfo{pages}{1} (\bibinfo{year}{2005}).

\bibitem[{\citenamefont{Kirchner and Si}(2010)}]{Kirchner2010}
\bibinfo{author}{\bibfnamefont{S.}~\bibnamefont{Kirchner}} \bibnamefont{and}
  \bibinfo{author}{\bibfnamefont{Q.}~\bibnamefont{Si}},
  \bibinfo{journal}{phys.~stat.~sol.~b} \textbf{\bibinfo{volume}{247}},
  \bibinfo{pages}{631} (\bibinfo{year}{2010}).

\bibitem[{\citenamefont{Caso et~al.}(2011)\citenamefont{Caso, Arrachea, and
  Lozano}}]{Caso2011}
\bibinfo{author}{\bibfnamefont{A.}~\bibnamefont{Caso}},
  \bibinfo{author}{\bibfnamefont{L.}~\bibnamefont{Arrachea}}, \bibnamefont{and}
  \bibinfo{author}{\bibfnamefont{G.~S.} \bibnamefont{Lozano}},
  \bibinfo{journal}{Phys.~Rev.~B} \textbf{\bibinfo{volume}{83}},
  \bibinfo{pages}{1} (\bibinfo{year}{2011}).

\bibitem[{\citenamefont{Gegenwart et~al.}(2008)\citenamefont{Gegenwart, Si, and
  Steglich}}]{Gegenwart.08}
\bibinfo{author}{\bibfnamefont{P.}~\bibnamefont{Gegenwart}},
  \bibinfo{author}{\bibfnamefont{Q.}~\bibnamefont{Si}}, \bibnamefont{and}
  \bibinfo{author}{\bibfnamefont{F.}~\bibnamefont{Steglich}},
  \bibinfo{journal}{Nat.~Phys.} \textbf{\bibinfo{volume}{4}},
  \bibinfo{pages}{186} (\bibinfo{year}{2008}).

\bibitem[{\citenamefont{Zhu et~al.}(2007)\citenamefont{Zhu, Kirchner, Bulla,
  and Si}}]{Zhu.06}
\bibinfo{author}{\bibfnamefont{J.}~\bibnamefont{Zhu}},
  \bibinfo{author}{\bibfnamefont{S.}~\bibnamefont{Kirchner}},
  \bibinfo{author}{\bibfnamefont{R.}~\bibnamefont{Bulla}}, \bibnamefont{and}
  \bibinfo{author}{\bibfnamefont{Q.}~\bibnamefont{Si}},
  \bibinfo{journal}{Phys.~Rev.~Lett.} \textbf{\bibinfo{volume}{99}},
  \bibinfo{pages}{227204} (\bibinfo{year}{2007}).

\bibitem[{\citenamefont{Vojta and Bulla}(2002)}]{Vojta.02}
\bibinfo{author}{\bibfnamefont{M.}~\bibnamefont{Vojta}} \bibnamefont{and}
  \bibinfo{author}{\bibfnamefont{R.}~\bibnamefont{Bulla}},
  \bibinfo{journal}{Phys.~Rev.~B} \textbf{\bibinfo{volume}{65}},
  \bibinfo{pages}{014511} (\bibinfo{year}{2002}).

\bibitem[{\citenamefont{Chen et~al.}(2011)\citenamefont{Chen, Li, Cullen,
  Williams, and Fuhrer}}]{Chen2011}
\bibinfo{author}{\bibfnamefont{J.-H.} \bibnamefont{Chen}},
  \bibinfo{author}{\bibfnamefont{L.}~\bibnamefont{Li}},
  \bibinfo{author}{\bibfnamefont{W.~G.} \bibnamefont{Cullen}},
  \bibinfo{author}{\bibfnamefont{E.~D.} \bibnamefont{Williams}},
  \bibnamefont{and} \bibinfo{author}{\bibfnamefont{M.~S.}
  \bibnamefont{Fuhrer}}, \bibinfo{journal}{Nature Physics}
  \textbf{\bibinfo{volume}{7}}, \bibinfo{pages}{535} (\bibinfo{year}{2011}),
  ISSN \bibinfo{issn}{1745-2473},
  \urlprefix\url{http://dx.doi.org/10.1038/nphys1962}.

\bibitem[{\citenamefont{Zhuravlev et~al.}(2007)\citenamefont{Zhuravlev,
  Zharekeshev, Gorelov, Lichtenstein, Mucciolo, and Kettemann}}]{Zhuravlev.07}
\bibinfo{author}{\bibfnamefont{A.}~\bibnamefont{Zhuravlev}},
  \bibinfo{author}{\bibfnamefont{I.}~\bibnamefont{Zharekeshev}},
  \bibinfo{author}{\bibfnamefont{E.}~\bibnamefont{Gorelov}},
  \bibinfo{author}{\bibfnamefont{A.~I.} \bibnamefont{Lichtenstein}},
  \bibinfo{author}{\bibfnamefont{E.~R.} \bibnamefont{Mucciolo}},
  \bibnamefont{and}
  \bibinfo{author}{\bibfnamefont{S.}~\bibnamefont{Kettemann}},
  \bibinfo{journal}{Phys.~Rev.~Lett.} \textbf{\bibinfo{volume}{99}},
  \bibinfo{pages}{247202} (\bibinfo{year}{2007}).

\bibitem[{\citenamefont{{Dias da Silva} et~al.}(2006)\citenamefont{{Dias da
  Silva}, Ingersent, Sandler, and Ulloa}}]{Diasdasilva.06}
\bibinfo{author}{\bibfnamefont{L.~G.} \bibnamefont{{Dias da Silva}}},
  \bibinfo{author}{\bibfnamefont{K.}~\bibnamefont{Ingersent}},
  \bibinfo{author}{\bibfnamefont{N.}~\bibnamefont{Sandler}}, \bibnamefont{and}
  \bibinfo{author}{\bibfnamefont{S.}~\bibnamefont{Ulloa}},
  \bibinfo{journal}{Phys.~Rev.~Lett.} \textbf{\bibinfo{volume}{97}},
  \bibinfo{pages}{096603} (\bibinfo{year}{2006}).

\bibitem[{\citenamefont{Withoff and Fradkin}(1990)}]{Withoff1990}
\bibinfo{author}{\bibfnamefont{D.}~\bibnamefont{Withoff}} \bibnamefont{and}
  \bibinfo{author}{\bibfnamefont{E.}~\bibnamefont{Fradkin}},
  \bibinfo{journal}{Phys.~Rev.~Lett.} \textbf{\bibinfo{volume}{64}},
  \bibinfo{pages}{1835} (\bibinfo{year}{1990}).

\bibitem[{\citenamefont{{}Bulla et~al.}(1997)\citenamefont{{}Bulla, {}Pruscke,
  and {}Hewson}}]{Bulla.97}
\bibinfo{author}{\bibfnamefont{R.}~\bibnamefont{{}Bulla}},
  \bibinfo{author}{\bibfnamefont{T.}~\bibnamefont{{}Pruscke}},
  \bibnamefont{and} \bibinfo{author}{\bibfnamefont{A.}~\bibnamefont{{}Hewson}},
  \bibinfo{journal}{J. Phys.: Condens. Matter} \textbf{\bibinfo{volume}{9}},
  \bibinfo{pages}{10463} (\bibinfo{year}{1997}).

\bibitem[{\citenamefont{Gonzalez-Buxton and Ingersent}(1998)}]{Buxton.98}
\bibinfo{author}{\bibfnamefont{C.}~\bibnamefont{Gonzalez-Buxton}}
  \bibnamefont{and}
  \bibinfo{author}{\bibfnamefont{K.}~\bibnamefont{Ingersent}},
  \bibinfo{journal}{Phys.~Rev.~B} \textbf{\bibinfo{volume}{57}},
  \bibinfo{pages}{14254} (\bibinfo{year}{1998}).

\bibitem[{\citenamefont{Logan and Glossop}(2000)}]{Logan.00}
\bibinfo{author}{\bibfnamefont{D.~E.} \bibnamefont{Logan}} \bibnamefont{and}
  \bibinfo{author}{\bibfnamefont{M.~T.} \bibnamefont{Glossop}},
  \bibinfo{journal}{J. Phys.: Condens. Matter} \textbf{\bibinfo{volume}{12}},
  \bibinfo{pages}{985} (\bibinfo{year}{2000}).

\bibitem[{\citenamefont{Vojta}(2001)}]{Vojta2001}
\bibinfo{author}{\bibfnamefont{M.}~\bibnamefont{Vojta}},
  \bibinfo{journal}{Phys.~Rev.~Lett.} \textbf{\bibinfo{volume}{87}},
  \bibinfo{pages}{097202} (\bibinfo{year}{2001}).

\bibitem[{\citenamefont{Ingersent and Si}(2002)}]{Ingersent.02}
\bibinfo{author}{\bibfnamefont{K.}~\bibnamefont{Ingersent}} \bibnamefont{and}
  \bibinfo{author}{\bibfnamefont{Q.}~\bibnamefont{Si}},
  \bibinfo{journal}{Phys.~Rev.~Lett.} \textbf{\bibinfo{volume}{89}},
  \bibinfo{pages}{076403} (\bibinfo{year}{2002}).

\bibitem[{\citenamefont{Glossop and Logan}(2003)}]{Glossop.03}
\bibinfo{author}{\bibfnamefont{M.~T.} \bibnamefont{Glossop}} \bibnamefont{and}
  \bibinfo{author}{\bibfnamefont{D.~E.} \bibnamefont{Logan}},
  \bibinfo{journal}{Europhys.~Lett.} \textbf{\bibinfo{volume}{61}},
  \bibinfo{pages}{810} (\bibinfo{year}{2003}).

\bibitem[{\citenamefont{Glossop et~al.}(2005)\citenamefont{Glossop, Jones, and
  Logan}}]{Glossop.05}
\bibinfo{author}{\bibfnamefont{M.~T.} \bibnamefont{Glossop}},
  \bibinfo{author}{\bibfnamefont{G.~E.} \bibnamefont{Jones}}, \bibnamefont{and}
  \bibinfo{author}{\bibfnamefont{D.~E.} \bibnamefont{Logan}},
  \bibinfo{journal}{J.~Phys.~Chem.~B} \textbf{\bibinfo{volume}{109}},
  \bibinfo{pages}{6564} (\bibinfo{year}{2005}).

\bibitem[{\citenamefont{Fritz et~al.}(2006)\citenamefont{Fritz, Florens, and
  Vojta}}]{Fritz2006}
\bibinfo{author}{\bibfnamefont{L.}~\bibnamefont{Fritz}},
  \bibinfo{author}{\bibfnamefont{S.}~\bibnamefont{Florens}}, \bibnamefont{and}
  \bibinfo{author}{\bibfnamefont{M.}~\bibnamefont{Vojta}},
  \bibinfo{journal}{Phys.~Rev.~B} \textbf{\bibinfo{volume}{74}},
  \bibinfo{pages}{144410} (\bibinfo{year}{2006}).

\bibitem[{\citenamefont{Glossop et~al.}(2011)\citenamefont{Glossop, Kirchner,
  Pixley, and Si}}]{Glossop.11}
\bibinfo{author}{\bibfnamefont{M.~T.} \bibnamefont{Glossop}},
  \bibinfo{author}{\bibfnamefont{S.}~\bibnamefont{Kirchner}},
  \bibinfo{author}{\bibfnamefont{J.}~\bibnamefont{Pixley}}, \bibnamefont{and}
  \bibinfo{author}{\bibfnamefont{Q.}~\bibnamefont{Si}},
  \bibinfo{journal}{Phys.~Rev.~Lett.} \textbf{\bibinfo{volume}{107}},
  \bibinfo{pages}{076404} (\bibinfo{year}{2011}).

\bibitem[{\citenamefont{Fritz and Vojta}(2013)}]{Fritz.13}
\bibinfo{author}{\bibfnamefont{L.}~\bibnamefont{Fritz}} \bibnamefont{and}
  \bibinfo{author}{\bibfnamefont{M.}~\bibnamefont{Vojta}},
  \bibinfo{journal}{Rep.~Prog.~Phys.} \textbf{\bibinfo{volume}{76}},
  \bibinfo{pages}{032501} (\bibinfo{year}{2013}).

\bibitem[{\citenamefont{Zamani et~al.}(2015)\citenamefont{Zamani, Ribeiro, and
  Kirchner}}]{Zamani}
\bibinfo{author}{\bibfnamefont{F.}~\bibnamefont{Zamani}},
  \bibinfo{author}{\bibfnamefont{P.}~\bibnamefont{Ribeiro}}, \bibnamefont{and}
  \bibinfo{author}{\bibfnamefont{S.}~\bibnamefont{Kirchner}},
  \bibinfo{journal}{(unpublished)}  (\bibinfo{year}{2015}).

\bibitem[{\citenamefont{Chung and Zhang}(2012)}]{Chung2012}
\bibinfo{author}{\bibfnamefont{C.-H.} \bibnamefont{Chung}} \bibnamefont{and}
  \bibinfo{author}{\bibfnamefont{K.~Y.-J.} \bibnamefont{Zhang}},
  \bibinfo{journal}{Phys.~Rev.~B} \textbf{\bibinfo{volume}{85}},
  \bibinfo{pages}{195106} (\bibinfo{year}{2012}).

\bibitem[{\citenamefont{Schir\'{o}}(2012)}]{Schiro2012a}
\bibinfo{author}{\bibfnamefont{M.}~\bibnamefont{Schir\'{o}}},
  \bibinfo{journal}{Phys.~Rev.~B} \textbf{\bibinfo{volume}{86}},
  \bibinfo{pages}{161101} (\bibinfo{year}{2012}).

\bibitem[{\citenamefont{Kaminski et~al.}(2000)\citenamefont{Kaminski, Nazarov,
  and Glazman}}]{Kaminski2000}
\bibinfo{author}{\bibfnamefont{A.}~\bibnamefont{Kaminski}},
  \bibinfo{author}{\bibfnamefont{Y.}~\bibnamefont{Nazarov}}, \bibnamefont{and}
  \bibinfo{author}{\bibfnamefont{L.~I.} \bibnamefont{Glazman}},
  \bibinfo{journal}{Physical Review B} \textbf{\bibinfo{volume}{62}},
  \bibinfo{pages}{8154} (\bibinfo{year}{2000}).

\bibitem[{\citenamefont{Parcollet et~al.}(1998)\citenamefont{Parcollet,
  Georges, Kotliar, and Sengupta}}]{Parcollet1998}
\bibinfo{author}{\bibfnamefont{O.}~\bibnamefont{Parcollet}},
  \bibinfo{author}{\bibfnamefont{A.}~\bibnamefont{Georges}},
  \bibinfo{author}{\bibfnamefont{G.}~\bibnamefont{Kotliar}}, \bibnamefont{and}
  \bibinfo{author}{\bibfnamefont{A.}~\bibnamefont{Sengupta}},
  \bibinfo{journal}{Phys.~Rev.~B} \textbf{\bibinfo{volume}{58}},
  \bibinfo{pages}{3794} (\bibinfo{year}{1998}).

\bibitem[{sup()}]{suppl}
\bibinfo{note}{See Supplementary Material}.

\bibitem[{\citenamefont{Werner and Eckstein}(2012)}]{Werner.12}
\bibinfo{author}{\bibfnamefont{P.}~\bibnamefont{Werner}} \bibnamefont{and}
  \bibinfo{author}{\bibfnamefont{M.}~\bibnamefont{Eckstein}},
  \bibinfo{journal}{Phys.~Rev.~B} \textbf{\bibinfo{volume}{86}},
  \bibinfo{pages}{045119} (\bibinfo{year}{2012}).

\bibitem[{\citenamefont{Foini et~al.}(2011)\citenamefont{Foini, Cugliandolo,
  and Gambassi}}]{Foini2011}
\bibinfo{author}{\bibfnamefont{L.}~\bibnamefont{Foini}},
  \bibinfo{author}{\bibfnamefont{L.}~\bibnamefont{Cugliandolo}},
  \bibnamefont{and} \bibinfo{author}{\bibfnamefont{A.}~\bibnamefont{Gambassi}},
  \bibinfo{journal}{Phys.~Rev.~B} \textbf{\bibinfo{volume}{84}},
  \bibinfo{pages}{1} (\bibinfo{year}{2011}).

\bibitem[{Note1()}]{Note1}
\bibinfo{note}{similar results hold for the Keldysh component of $\chi
  $.}

\bibitem[{\citenamefont{Sonner and Green}(2012)}]{Sonner.12}
\bibinfo{author}{\bibfnamefont{J.}~\bibnamefont{Sonner}} \bibnamefont{and}
  \bibinfo{author}{\bibfnamefont{A.~G.} \bibnamefont{Green}},
  \bibinfo{journal}{Phys.~Rev.~Lett} \textbf{\bibinfo{volume}{109}},
  \bibinfo{pages}{091601} (\bibinfo{year}{2012}).

\bibitem[{\citenamefont{Bhaseen et~al.}(2015)\citenamefont{Bhaseen, Doyon,
  Lucas, and Schalm}}]{Bhaseen.15}
\bibinfo{author}{\bibfnamefont{M.~J.} \bibnamefont{Bhaseen}},
  \bibinfo{author}{\bibfnamefont{B.}~\bibnamefont{Doyon}},
  \bibinfo{author}{\bibfnamefont{A.}~\bibnamefont{Lucas}}, \bibnamefont{and}
  \bibinfo{author}{\bibfnamefont{K.}~\bibnamefont{Schalm}},
  \bibinfo{journal}{Nature Phys.} \textbf{\bibinfo{volume}{11}},
  \bibinfo{pages}{509} (\bibinfo{year}{2015}).

\end{thebibliography}
\end{document}